\documentclass[twoside]{IEEEtran}
\usepackage{cite} 
\usepackage{wrapfig,graphicx,booktabs,fancyhdr,amsfonts}
\usepackage{bm,amssymb,amsmath,amsthm,wasysym,multirow,dsfont}
\usepackage{subcaption,algorithm,xcolor,pifont,float,enumitem}
\usepackage[noend]{algorithmic}

\newcommand{\ba}{\begin{array}}
\newcommand{\ea}{\end{array}}

\newcommand{\sinc}{{\rm sinc}}
\renewcommand{\equiv}{\triangleq}
\DeclareMathAlphabet{\mathpzc}{OT1}{pzc}{m}{it}

\begin{document}
\title{GNSS Signal Authentication via Power and Distortion Monitoring}
\author{Kyle~D.~Wesson,~\IEEEmembership{Member,~IEEE,}~Jason~N.~Gross,~\IEEEmembership{Member,~IEEE,}
  ~Todd~E.~Humphreys,~\IEEEmembership{Member,~IEEE,} and
  Brian~L.~Evans,~\IEEEmembership{Fellow, IEEE}%
  \thanks{Manuscript submitted on DD MMM 2017}%
  \thanks{K.~Wesson was supported in part by DoD, Air Force Office of
    Scientific Research, National Defense Science and Engineering Graduate
    Fellowship 32 CFR $\S$168a, and The University of Texas at Austin
    Microelectronics and Computer Development Fellowship. J. ~Gross was
    supported in part by a West Virginia University Big XII Faculty
    Fellowship. T. Humphreys was supported in part by the National Science
    Foundation under Grant No. 1454474 (CAREER).}%
  \thanks{K. ~Wesson, B. ~Evans are with the Department of Electrical and
    Computer Engineering, and T. Humphreys with the Department of Aerospace
    Engineering and Engineering Mechanics, at The University of Texas at
    Austin, Austin, TX 78712 USA (e-mail: kyle.wesson@utexas.edu,
    todd.humphreys@mail.utexas.edu, bevans@ece.utexas.edu). J. ~Gross is with
    the Department of Mechanical and Aerospace Engineering at West Virginia
    University (e-mail: jason.gross@mail.wvu.edu).} }

\maketitle


\begin{abstract}
  We propose a simple low-cost technique that enables civil Global Positioning
  System (GPS) receivers and other civil global navigation satellite system
  (GNSS) receivers to reliably detect carry-off spoofing and jamming. The
  technique, which we call the Power-Distortion detector, classifies received
  signals as interference-free, multipath-afflicted, spoofed, or jammed
  according to observations of received power and correlation function
  distortion.  It does not depend on external hardware or a network connection
  and can be readily implemented on many receivers via a firmware
  update. Crucially, the detector can with high probability distinguish
  low-power spoofing from ordinary multipath. In testing against over 25
  high-quality empirical data sets yielding over 900,000 separate detection
  tests, the detector correctly alarms on all malicious spoofing or jamming
  attacks while maintaining a $<$0.6\% single-channel false alarm rate.
\end{abstract}


\section{Introduction}
\IEEEPARstart{G}{NSS} receivers are tremendously popular in navigation and
timing applications due to their accuracy, low cost, and global operation.
Low cost can be credited to the fact that civil GNSS signals are
defined in freely-available, open-access standards~\cite{isGps200Latest,
  icdGalileoOsLatest}, which makes receiver development straightforward.  But
an open-access standard, together with civil GNSS signals' near-perfect
predictability, invites forgery: receivers can fall victim to spoofing attacks
in which counterfeit GNSS signals fool the receiver into reporting a
hazardously misleading position or time~\cite{t_humphreys_gcs08}. The
vulnerability of civil GNSS receivers to spoofing is a serious risk for
GNSS-dependent critical infrastructure and safety-of-life
applications~\cite{volpe2001gps, shepard2012_IJCIP}.

GNSS authentication techniques can be broadly categorized into three groups: (1)
cryptographic techniques that exploit unpredictable but verifiable signal
modulation in the GNSS spreading code or navigation data, (2) geometric
techniques that exploit the angle-of-arrival diversity of authentic GNSS
signals, and (3) GNSS signal processing techniques that do not fall into
categories (1) or (2). A comprehensive survey of GNSS authentication
techniques is offered in~\cite{psiaki2016gnssSpoofing}.

Cryptographic signal authentication is effective ~\cite{psiaki2016gnssSpoofing,
  wesson2011cga, humphreys2011ds}, but, despite recent interest and engagement
by U.S. and European satellite navigation agencies~\cite{kerns2014nmaimp,
  fernandez2016navigation}, no open civil GNSS signals yet incorporate
cryptographic modulation.  Moreover, it has become clear that financial and
technical hurdles will impede development and implementation of such
modulation for years to come.  It is possible to leverage the existing
encryption of military GNSS signals for civil signal
authentication~\cite{psiaki2012_dualRxCorr}, but this technique requires the
protected receiver to be connected to a data network, an undesirable
dependency that prevents stand-alone operation.

Authentication techniques that exploit GNSS signals' geometric diversity can
also be highly effective.  These include angle-of-arrival discrimination
techniques based on multiple antennas~\cite{montgomery2009pcgs,
  psiaki2014gnss, meurer2012robust,borio2013panova}, or a single antenna
experiencing oscillatory motion~\cite{psiaki2013motionION}, or a single
antenna with multiple feeds~\cite{mcmilin2014single}.  The drawback of these
approaches is their reliance on multiple antennas, antenna motion, or an
assumption that interference signals arrive from below the antenna's horizon.
Likewise, methods that require coupling with inertial sensors
\cite{white1998detection, tanil2015gnss} or vision sensors may prove
impractical in applications with cost, size, weight, or power constraints.

Practical near-term GNSS signal authentication techniques are those that do
not require changes to GNSS signals-in-space, are receiver-autonomous,
low-cost, require no additional hardware, and can be implemented via a
software or firmware update. Recognizing the value of techniques that fall
into this category, previous work has proposed monitoring the total received
power via the Automatic Gain Control (AGC) setpoint~\cite{akos2012s}, and
monitoring autocorrelation profile distortion~\cite{ledvina2010cgr,
  cavaleri2010dscc, huang2016gnss}.  But when acting separately these
techniques are unreliable for signal authentication. A received power monitor
that ignores correlation distortion may not detect a low-power
spoofer. Moreover, because a power-monitoring-only technique does not
distinguish between spoofing and jamming, its alarm rate can become
intolerable in urban areas where so-called personal privacy devices (PPDs,
small GNSS jammers)~\cite{mitch2012kyejam} are common.  For their part, the
distortion-monitoring approaches in \cite{ledvina2010cgr, cavaleri2010dscc,
  huang2016gnss} ignore total received power, and thus can be fooled by a
spoofer transmitting with a significant power advantage over the authentic
signals, which, by action of the AGC, forces the authentic signals under the
noise floor, leaving a distortion-free correlation
function~\cite{humphreys2011ds}.

A GNSS authentication technique is needed that is both practical in the sense
described above and reliable at detecting spoofing.  To this end, we propose
to combine the elemental tests mentioned earlier, namely, detection of
anomalous received power and detection of correlation profile distortion, in a
GNSS signal authentication technique we call the Power-Distortion detector, or
PD detector for short.  The key insight behind our approach is this: The
practically-unavoidable interaction between authentic and spoofed GNSS signals
during the initial stages of a tracking-points-carry-off spoofing attack makes
such spoofing evident, with high probability, in received power or signal
distortion or both. If a carry-off-type spoofer transmits at a low signal
power, the attack will either be ineffective or will cause significant
correlation function distortion as the similarly-sized spoofing and authentic
signals interact. On the other hand, if a spoofer transmits at a high signal
power, the correlation function may be distortion-free but the receiver's
total received power will be anomalously high.  Our proposed technique traps a
would-be spoofer between simultaneous measurements of received power and
correlation function distortion.  Provided the spoofer is unable to block or
otherwise null the authentic GNSS signals impinging on the receiver's antenna
(a difficult task if the receiver enjoys a physical security
perimeter~\cite{psiaki2016gnssSpoofing}), the combination of these
measurements within a Bayesian detection framework poses a formidable defense
against carry-off-type spoofing.

This paper, a significant extension of our work in~\cite{wesson2013sandwich},
makes three main contributions. First, it introduces a novel technique for
detecting GNSS jamming and carry-off-type spoofing and rigorously develops the
measurement models and probability distributions required to characterize the
detection statistic.  Second, it presents a Monte-Carlo-type method for
determining the Bayes-optimal decision rule and offers detailed consideration
of the requisite cost function.  Third, it presents a thorough evaluation of
the proposed technique against three realistic data sets: (1) the Texas
Spoofing Test
Battery~\cite{humphreys2012_TEST_Battery,rnlTexbatSite,texbat_ds7_ds8}, a
public set of GPS spoofing recordings; (2) the RNL Multipath and Interference
Recordings~\cite{wesson2011vsd, pesyna2011tightly}, a public set of deep urban
GNSS recordings with significant multipath; and (3) a set of recordings of
GNSS signals subject to jamming.

\section{Signal Models}
\label{sec:sign-interf-model}

\subsection{Pre-Correlation Model}
Consider the following generic representation of an authentic GNSS signal
exiting a receiver's radio frequency (RF) front-end downconversion chain.  For
notational compactness, the signal is expressed by its complex baseband
representation,
\begin{equation}
  \label{eq:rA}
  r_{\rm A}(t) = \sqrt{P_{\rm A}}D(t - \tau_{\rm A})C_r(t - \tau_{\rm A}) \exp(j
     \theta_{\rm A})  
\end{equation}
where $t$ is time in seconds, $P_{\rm A}$ is the received power of the
authentic signal in Watts, $D(t)$ is the $\pm 1$-valued navigation data
modulation, $C_r(t)$ is the $\pm 1$-valued pseudorandom spreading (ranging)
code, $\tau_{\rm A}$ is the code phase in seconds, and
$\exp(j \theta_{\rm A})$ is the carrier with phase $\theta_{\rm A}$ in
radians.  $P_{\rm A}$, $\tau_{\rm A}$, and $\theta_{\rm A}$ are assumed to
vary with time; their time dependency is suppressed for notational
simplicity.  Without loss of generality for the purposes of this paper, the
navigation data modulation $D(t)$ can be ignored; hence, hereafter we assume
$D(t) = 1$.

Let $r_{\rm I}(t)$ represent a single complex-valued interference signal that
is structurally identical to $r_{\rm A}(t)$.  This could be a multipath,
spoofing, or jamming signal. If multipath, $r_{\rm I}(t)$ represents the
strongest reflection at time $t$, whose effect on received power and
correlation function distortion is a good proxy for that of the aggregate
multipath.  If jamming, $r_{\rm I}(t)$ represents structured-signal jamming,
an especially potent form of jamming similar to spoofing except there is no
expected correlation between the jamming signal's code or carrier phase and
those of $r_{\rm A}(t)$ \cite{humphreysGNSShandbook}.  The interference signal
is modeled as
\begin{align}
  \label{eq:si}
  r_{\rm I}(t) = \sqrt{\eta P_{\rm A}}C_r(t-\tau_{\rm I})\exp(j \theta_{\rm I})
\end{align}
where $\eta = P_{\rm I}/P_{\rm A}$ is the interference signal's power
advantage relative to the authentic signal, and $\tau_{\rm I}$ and
$\theta_{\rm I}$ are the interference signal's code and carrier phase,
respectively.  

To complete the received signal model, let
\begin{equation}
  \label{eq:rN_model}
  r_{\rm N}(t) = N(t) + M(t)
\end{equation}
be a white zero-mean complex-valued Gaussian process that models the sum of
thermal noise $N(t)$, with constant spectral density $N_0$, and multi-access
interference $M(t)$, with variable spectral density $M_0$. The two noise
components are assumed to be independent so that $r_{\rm N}(t)$ has density
$N_0 + M_0$.  $M(t)$ accounts for the noise contribution from other legitimate
GNSS signals besides the desired signal (collectively called multi-access
signals), and any interference accompanying these.  With the addition of
$r_{\rm N}(t)$, the full received signal-plus-interference-and-noise model is
given by
\begin{equation}
  \label{eq:rt_model}
r(t) = r_{\rm A}(t) + r_{\rm I}(t) + r_{\rm N}(t)  
\end{equation}

\begin{figure*}[htb!]
  \centering
  \includegraphics[width=\textwidth]{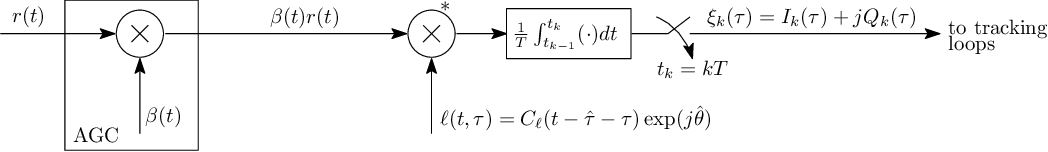}
  \caption{Block diagram of the standard AGC, correlation, and accumulation
    operations in a GNSS receiver. The product of the AGC-scaled incoming
    signal $\beta(t)r(t)$ and the conjugate of the local replica
    $\ell(t,\tau)$ is accumulated over $T$ seconds to produce the discrete
    complex-valued accumulation product $\xi_k(\tau)$. For notational
    convenience, the accumulation product has been scaled by $1/T$.}
  \label{fig:rx-model}
\end{figure*}

As shown in Fig.~\ref{fig:rx-model}, an AGC circuit is assumed to apply a
scaling factor $\beta(t)$ to $r(t)$ so that the power in the scaled signal
$\beta(t)r(t)$ remains constant.  Subsequent to AGC scaling, the signal is
quantized and encoded; for simplicity, these operations are ignored in
Fig. \ref{fig:rx-model} and in the remainder of this paper, as their effects
on the this paper's detection processing are negligible.

At the core of GNSS signal processing is correlation of $\beta(t)r(t)$ with a
local replica
\[\ell(t,\tau) = C_{\ell}(t - \hat{\tau} - \tau) \exp(j \hat{\theta})\] where
$\tau$ is an arbitrary code phase lag in seconds and $C_{\ell}(t)$ is the
local code replica, which, ignoring the effects of band-limiting on the
received signal, is often made equal to $C_r(t)$. The goal of a receiver's
code and carrier tracking loops is to drive the estimates $\hat{\tau}$ and
$\hat{\theta}$ to match $\tau_{\rm A}$ and $\theta_{\rm A}$ as accurately as
possible.  In practice, however, $\hat{\tau}$ and $\hat{\theta}$ track the
code and carrier phase of the composite signal $r(t)$, not just those of
$r_{\rm A}(t)$.

\subsection{Post-Correlation Model}
\begin{figure}[htb]
  \centering
  \includegraphics[width=0.48\textwidth,trim={0 0 0 0.2cm},clip]
  {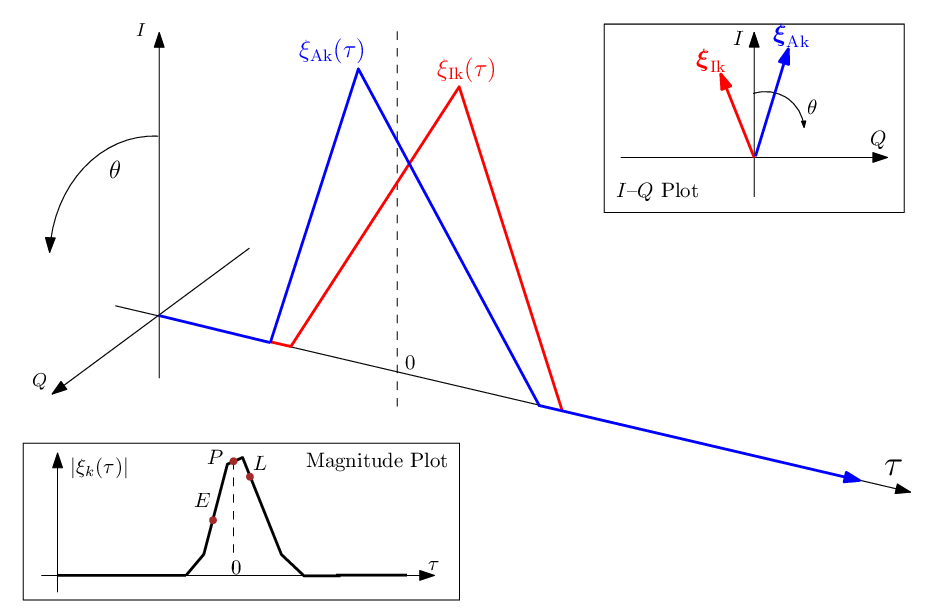}
  \caption{Components of the correlation function $\xi_k(\tau)$ for an example
    scenario with a triangular-shaped $R(\tau)$ and a strong spoofing or
    multipath interference component $\xi_{{\rm I}k}(\tau)$.  For visual
    clarity, the thermal noise component $\xi_{{\rm N}k}(\tau)$ is not shown.
    The upper-right inset shows the phase angle of the authentic and
    interference components relative to the in-phase $I$ and quadrature $Q$
    components of the local replica $\ell(t,\tau)$.  The lower-left inset
    shows the magnitude of the composite correlation function $\xi_k(\tau)$,
    which is clearly distorted by the interference component.  $E$, $P$, and
    $L$ mark the early, prompt, and late correlation taps, respectively.  The
    local replica's code phase estimate $\hat{\tau}$ attempts to track the
    code phase of the composite signal $r(t)$ by equalizing $E$ and $L$.  The
    prompt tap $P$ is located at $\tau = 0$.}
  \label{fig:phasor_relation}
\end{figure}

Correlation and accumulation over an interval $T$ ending at time $t_k = kT$,
$k \in \{1, 2,...\}$ produce the complex-valued accumulation product $\xi_k$,
which, when viewed as a function of the arbitrary lag $\tau$ introduced by the
local replica $\ell(t,\tau)$, is called the receiver's correlation function
for signal $r_{\rm A}(t)$ at time $t_k$, and is modeled as
\cite{vannee1993ssme}
\begin{align}
  \label{eq:af}
  \xi_k(\tau) = \beta_k [ \xi_{{\rm A}k}(\tau) + \xi_{{\rm I}k}(\tau) + \xi_{{\rm N}k}(\tau) ]
\end{align}
where $\beta_k$ is the average value of $\beta(t)$ over the $k$th accumulation
interval, and $\xi_{{\rm A}k}(\tau)$, $\xi_{{\rm I}k}(\tau)$,
$\xi_{{\rm N}k}(\tau)$ are the complex correlation function components
corresponding to the authentic signal, the interference signal, and thermal
noise, respectively.  Fig.~\ref{fig:phasor_relation} illustrates such a
correlation function.

The function $R(\tau) = E[C_r(t)C_{\ell}(t - \tau)]$, often called the
autocorrelation function of $C_r(t)$ even though, strictly speaking,
$C_{\ell}(t)$ may be slightly different from $C_r(t)$, approximates the
interaction between $C_r(t)$ and $C_{\ell}(t)$ over the correlation and
accumulation operations:
\begin{equation}
  \nonumber
  R(\tau) \approx \frac{1}{T}\int_{t_{k-1}}^{t_k} C_r(t) C_\ell(t - \tau) dt
\end{equation}
The correlation components $\xi_{{\rm A}k}(\tau)$ and $\xi_{{\rm I}k}(\tau) $
can be modeled in terms of $R(\tau)$ as 
\begin{align*}
  \xi_{{\rm A}k}(\tau) & = \sqrt{P_{{\rm A}k}} R(-\Delta \tau_{{\rm A}k} + \tau) \exp(j
  \Delta \theta_{{\rm A}k}) \\
  \xi_{{\rm I}k}(\tau) & = \sqrt{\eta_kP_{{\rm A}k}}
  R(-\Delta \tau_{{\rm I}k} + \tau) \exp(j \Delta \theta_{{\rm I}k})
\end{align*}
where $P_{{\rm A}k}$ and $\eta_k$ are the average values of $P_{{\rm A}}$ and
$\eta$ over the accumulation interval, and $\Delta \tau_{{\rm A}k}$ is the
average value of the difference $\tau_{\rm A} - \hat{\tau}$ over the
accumulation interval, with similar definitions for $\Delta \tau_{{\rm I}k}$,
$\Delta \theta_{{\rm A}k}$, and $\Delta \theta_{{\rm I}k}$.

The thermal noise component of the correlation function, $\xi_{{\rm
    N}k}(\tau)$, is modeled as having independent in-phase (real) and
quadrature (imaginary) components, each modeled as a zero-mean Gaussian white
discrete-time process:
\begin{align*}
  E[\mathds{R}\{\xi_{{\rm N}k}(\rho)\} \mathds{I} \{\xi_{{\rm N}j}(\nu)\}] = 0 \quad \forall
  ~\rho,\nu,k \neq j
\end{align*}
The chip interval of the spreading code $C_r(t)$, denoted $\tau_c$, ranges from
0.01 to 1 $\mu$s in modern GNSS signals.  Due to the pseudorandom nature of
$C_r(t)$, only samples of $\xi_{{\rm N}k}(\tau)$ within $2\tau_c$ of each other
are correlated \cite{dierendonck1992ncs}:
\begin{align*}
  E[\xi_{{\rm N}k}(\rho) \xi^*_{{\rm N}k}(\nu)] = \left\{\ba{ll} 2\sigma_{\rm N}^2(1-
  |\rho-\nu|/\tau_c) & |\rho-\nu|\leq 2\tau_c\\
   0 &  |\rho-\nu| > 2\tau_c \ea \right.
\end{align*}
Here, $^*$ denotes the complex conjugate and $\sigma_{\rm N}^2$ is the variance
of the in-phase and quadrature components of $\xi_{{\rm N}k}(\tau)$, which is
related to the spectral density of the white noise process $r_{\rm N}(t)$ by
$\sigma_{\rm N}^2 = (N_0 + M_0)/2T$.  

\section{Hypothesis Testing Framework}
\label{sec:mult-hypoth-test}
We adopt a Bayesian M-ary hypothesis testing framework
\cite[Ch. 2]{vPoor1994dae} for distinguishing between hypotheses
$H_i,~ i\in \mathcal{I}$, where $\mathcal{I} = \{0,1,2,3\}$. The null
hypothesis $H_0$ corresponds to the interference-free case, and
$H_i,~i = 1,2,3$ correspond respectively to multipath, spoofing, and jamming.

The foregoing signal models reveal three parameters relevant to choosing
between hypotheses, namely, the interference power advantage $\eta$, and the
interference-to-authentic code and carrier offsets
$\Delta \tau \equiv \tau_{\rm I} - \tau_{\rm A}$ and
$\Delta \theta \equiv \theta_{\rm I} - \theta_{\rm A}$.  Let these be combined
into a single vector ${\theta} = [\eta, \Delta \tau, \Delta \theta]^T$ assumed
to lie in the parameter space $\Lambda$, itself divided into disjoint
parameter sets $\Lambda_i \subset \Lambda$, $i\in\mathcal{I}$, each associated
with its corresponding hypothesis $H_i$.  Thus, deciding that
${\theta} \in \Lambda_i$ is equivalent to choosing hypothesis $H_i$.  Note
that, because ${\theta}$ can take on a range of values, the hypothesis testing
problem is composite.

In a Bayesian formulation of the composite hypothesis testing problem, the
parameter vector ${\theta}$ is viewed as a random quantity, ${\Theta}$, having
density $w({\theta})$, with $\pi_i \equiv P({\Theta} \in \Lambda_i)$ being the
prior probability that ${\Theta}$ falls in $\Lambda_i$.  We denote by
$w_i({\theta})$ the conditional density of ${\Theta}$ given that
${\Theta} \in \Lambda_i$; it follows that
\[ w_i({\theta}) = \left\{ \ba{ll} 0 & {\theta} \notin \Lambda_i \\
    w({\theta})/\pi_i & {\theta} \in \Lambda_i \ea \right. \]

We propose to decide between the four hypotheses based on two types of
observation at each $t_k$, namely, the received power measurement $P_k$ and
the symmetric difference measurement $D_k$, both detailed in a later
section. The observation vector ${z}_k = [D_k, P_k]^T$, which resides in the
observation set $\Gamma$, is modeled as a random variable ${Z}_k$ with
conditional density $p({z}_k | {\theta})$.  $H_i$ can be defined as the
hypothesis that ${Z}_k$ is distributed as
$p({z}_k | {\Theta} \in \Lambda_i), ~i\in\mathcal{I}$.

A decision rule $\delta({z}_k)$ is a partition of $\Gamma$ into disjoint
decision regions $\Gamma_i, ~i\in\mathcal{I}$, such that $H_i$ is chosen when
${z}_k \in \Gamma_i$:
\begin{equation*}
  \delta({z}_k) = \left\{\ba{ll} 
    0 &  \mbox{if} ~ {z}_k \in \Gamma_0 \\
    1 &  \mbox{if} ~{z}_k \in \Gamma_1 \\
    2 &  \mbox{if} ~{z}_k \in \Gamma_2 \\
    3 &  \mbox{if} ~{z}_k \in \Gamma_3 \ea \right.
\end{equation*}

Let $C[i,{\theta}]$ be the cost of choosing $H_i$ when ${\theta} \in \Lambda$
is the actual parameter vector.  Note that this function is sensitive to a
particular value of $\theta$, which makes it more general than one that simply
assigns a unform cost for choosing $H_i$ when $\theta \in \Lambda_j$.  A later
section will introduce various embodiments of $C[i,{\theta}]$.

An optimum rule selects the least costly hypothesis, on average, given the
observation ${z}_k$.  More precisely, if we define the conditional risk, or
the average cost for ${\Theta} = {\theta}$, as
\begin{equation*}
  R_{{\theta}}(\delta) \equiv E_{{\theta}}\left\{
    C[\delta({Z}_k),{\theta} ] \right\}, \quad {\theta} \in \Lambda
\end{equation*}
where $E_{{\theta}}$ denotes expectation assuming
${Z}_k \sim p({z}_k |\theta)$, and if we define average, or Bayes, risk as
\begin{equation}
  \label{eq:bayesOrAverageRisk}
  r(\delta) = E[R_{{\Theta}}(\delta)]
\end{equation}
where the expectation is now taken over the random quantity ${\Theta}$,
then the optimum rule $\delta$ is the one whose decision regions
$\Gamma_i, ~ i\in\mathcal{I}$, minimize $r(\delta)$.

To find the minimizing $\delta$, each parameter set $\Lambda_i$ and
conditional distribution $w_i({\theta})$ must be defined for
$i\in\mathcal{I}$.  These could be approximated from an extensive campaign of
empirical multipath, spoofing, and jamming data collection and analysis.  But
any empirical characterization runs the risk of biasing the $\Lambda_i$ and
$w_i({\theta})$ toward the particular dataset used.  This is merely a
restatement of Hume's problem of induction: metaphorically speaking, the
empirical dataset may only contain white swans \cite{hume2000enquiry}.  To
avoid this pitfall of inference, which, in a security context, could be a
serious vulnerability, the following definitions of $\Lambda_i$ and
$w_i({\theta})$ are informed not only by observation but also by the physical
characteristics and limitations of signals under the various hypotheses.

\subsection*{$H_0$: No interference}
In what follows, we denote the marginal distributions of $w_i({\theta})$ by
$w_{\eta i}(x)$, $w_{\Delta \tau i}(x)$, and
$w_{\Delta \theta i}(x), ~i\in\mathcal{I}$.  Under the null hypothesis
($i =0$), which corresponds to the interference-free case, we have
\[ \Lambda_0 = \{{\theta} \in \Lambda ~|~ \eta = 0 \}\] It follows that
$w_{\eta 0}(x)$ is equivalent to the Dirac delta function.  The marginal
distributions $w_{\Delta \tau 0}(x)$, and $w_{\Delta \theta 0}(x)$ can be
defined arbitrarily, since for $\eta = 0$ they have no effect on $r(\delta)$.

\subsection*{$H_1$: Multipath} In the case of multipath, $\eta$ and
$\Delta \tau$ can be bounded as follows:
\begin{align*}
 \Lambda_1 = \{ {\theta} \in \Lambda ~|~ 0 < \eta <
   \eta_1, 0 < \Delta \tau < \Delta \tau_1  \}
\end{align*}
The bounds $\eta_1$ and $\Delta \tau_1$ are informed by the physics of signal
reception and signal processing, as follows.  Whenever the authentic signal is
unobstructed, it arrives with greater power than any echo, whose additional
path length and interaction with reflection surfaces invariably attenuate its
power \cite{g_turin1972ump}.  Moreover, an echo whose delay is more than
double the spreading code chip interval (e.g., more than 0.2 $\mu$s for
$\tau_c = 0.1$ $\mu$s) causes no correlation function distortion
\cite{dierendonck1992ncs}, and, due to the additional path loss, is at least 3
dB weaker than an unobstructed authentic signal
\cite{steingass2004measuring}. Thus, its effect on $r(\delta)$ is negligible.
It follows that $\eta_1 = 1$ and $\Delta \tau_1 = 2\tau_c$ upper-bound the
multipath parameter set for an unobstructed authentic signal.

On the other hand, the severe shadowing experienced by mobile receivers can
cause $\eta > 1$, especially in urban environments where surrounding buildings
simultaneously attenuate and reflect authentic signals
\cite{steingass2004measuring}.  In these situations, there is no practical
upper limit on $\eta$ because the direct-path authentic signal may be
attenuated by 50 dB or more. It is not possible to reliably distinguish
multipath from low-power spoofing in such circumstances.

One can avoid this difficulty by applying this paper's detection test only
when the authentic signal $r_{\rm A}(t)$ is received without severe shadowing,
in which case the probability that $\eta > 1$ under $H_1$ again becomes
negligible. In particular, the multipath model developed below assumes that
$r_{\rm A}(t)$ is attenuated less than 6 dB by shadowing.  In practice, one
simply excludes cases in which the received power $P_k$ is not unusually high
yet the measured carrier-to-noise ratio $C/N_0$ drops by more than 6 dB from
its modeled value for an unobstructed authentic signal.

An analysis of GNSS multipath was carried out to validate the bounds
$\eta_1 = 1$ and $\Delta \tau_1 = 2\tau_c$ and to characterize
$w_1({\theta})$.  The analysis was based on the Land Mobile Satellite Channel
Model (LMSCM) \cite{lehner2007multipath}, itself based on extensive
experimentation with a wideband airborne transmitter at GNSS frequencies in
urban and suburban environments
\cite{steingass2004measuring,lehner2007multipath}.  The LMSCM generates power,
delay, and carrier phase for both line-of-sight and echo signals using
deterministic models for attenuation, diffraction, and delay that respond to
stochastically-generated obstacles and reflectors in the simulation
environment.

In keeping with the philosophy of robust detection,
multipath scenarios were sought whose distribution $w_1({\theta})$ was most
similar to the distribution for spoofing, $w_2({\theta})$.  From the
standpoint of distinguishing multipath from spoofing, this is the worst-case
$w_1({\theta})$.  As a practical matter, the worst-case $w_1({\theta})$ has a
high proportion of large $\eta$ values and a wide range of $\Delta \theta$.
Not surprisingly, urban LMSCM scenarios with low-elevation satellite signals
yielded the worst-case $w_1({\theta})$.

For flexibility of detector design, the simulation-derived $w_1({\theta})$ was
parameterized in terms of satellite elevation angle $\alpha_e$.  Several
1-minute LMSCM simulations with randomly-created urban environments and
various satellite azimuth angles were run for $\alpha_e \in [20,80]$ degrees.
At each simulation epoch, the maximum-amplitude echo was identified and its
relative power, delay, and phase with respect to the line-of-sight signal were
taken as sampled $\eta$, $\Delta \tau$, and $\Delta \theta$ values from which
$w_1({\theta})$ could be approximated.  Epochs whose line-of-sight signal was
attenuated by more than 6 dB were excluded, resulting in exclusion of up to
90\% of epochs for heavily-shadowed scenarios at $\alpha_e = 20$, but less
than 1\% of epochs at $\alpha_e = 80$.  At each $\alpha_e$, the scenario
producing the worst-case $w_1({\theta})$ was selected.

A statistical analysis of the sampled $\eta$, $\Delta \tau$, and
$\Delta \theta$ values revealed that the relative phase $\Delta \theta$ was
uniformly distributed on $[0, 2\pi)$ and independent of $\eta$, $\Delta \tau$,
which is unsurprising given the wavelength-scale sensitivity of
$\Delta \theta$ to path length.  The parameters $\eta$ and $\Delta \tau$ were
found to be significantly correlated, with a linear correlation coefficient of
approximately $\rho = -0.23$, implying that more distant reflectors tend to
produce weaker echos.  Consistent with \cite{g_turin1972ump}, the marginal
distribution $w_{\eta 1}(x)$ was found to be log-normal, with a mean of $-21$ dB
and a standard deviation of 5 dB, both insensitive to $\alpha_e$.  Thus,
excluding cases of heavy line-of-sight signal shadowing, multipath causes
$\eta > 1$ with negligible probability, which indicates that $\eta_1 = 1$ is
an unproblematic bound.  The marginal distribution $w_{\Delta \tau 1}(x)$ was
found to be exponential,
\begin{equation}
  \label{eq:exponential}
  w_{\Delta \tau 1}(x) = \mu^{-1} e^{-x/\mu}, \quad x \geq 0 \nonumber
\end{equation}
with $\mu$ a quadratic function of $\alpha_e$, 
\[ \mu = 0.012\alpha_e^2 -2.4\alpha_e + 134 \] where $\alpha_e$ is expressed
in degrees and $\Delta \tau$ in nanoseconds. Thus, $\Delta \tau$ values are
widely spread at low $\alpha_e$ but tightly clustered near zero at high
$\alpha_e$, consistent with the empirical distributions in
\cite{steingass2004measuring}.

\subsection*{$H_2$: Spoofing}
Under the spoofing hypothesis, we take $\eta_1 = 1$ as a lower bound for
$\eta$ and assume $|\Delta \tau|$ is upper-bounded by
$\Delta \tau_1 = 2\tau_c$:
\begin{align*}
  \Lambda_2 = \{ {\theta} \in \Lambda ~|~ \eta_1 \leq \eta, 
  0 < |\Delta \tau | < \Delta \tau_1 \}
\end{align*}
These bounds make sense because the spoofing signal must be at least as
powerful as the authentic signal for reliable spoofing, and because spoofing
whose $|\Delta \tau|$ value exceeds $\Delta \tau_1 = 2\tau_c$ is no longer
classified as carry-off-type spoofing, since the interference is uncorrelated
with the authentic signal.  Instead, for $|\Delta \tau| \geq \Delta \tau_1$,
the attack is classified as jamming.

For maximum stealth, the distribution $w_2({\theta})$ should be as close as
possible to $w_1({\theta})$ while respecting the bounds $ \eta_1$ and
$\Delta \tau_1$ and allowing a wide range of $\Delta \tau$ for code pull-off.
Thus, $w_{\eta 2}(x)$ is modeled as log-normal with mean of 1 dB and standard
deviation a fraction of a dB, $w_{\Delta \tau 2}(x)$ as exponential with
parameter $\mu = 120$ (mimicking low-elevation multipath), and
$w_{\Delta \theta 2}(x)$ as uniform on $[0, 2\pi)$.

\subsection*{$H_3$: Jamming}
Because it is uncorrelated with the authentic signal, jamming is taken to have
$|\Delta \tau| \geq \Delta \tau_1 = 2\tau_c$.  Its power advantage is assumed to be
at least $\eta_1 = 1$, as weaker jamming is both harmless and so common as to
be unremarkable.  Thus, the jamming parameter set is
\begin{align*}
  \Lambda_3 = \{ {\theta} \in \Lambda ~|~ \eta_1 \leq \eta, 
  \Delta \tau_1 \leq |\Delta \tau| \}
\end{align*}
The marginal distribution $w_{\eta 3}(x)$ is taken to be Rician.  By adjusting
the Rician distance and scale parameters $\nu$ and $\sigma$, one can model
high- or low-power jamming within a wide or narrow range.  The marginal
distributions $w_{\Delta \tau 3}(x)$ and $w_{\Delta \theta 3}(x)$ can be
modeled arbitrarily, as they have no effect on $r(\delta)$ since
$|\Delta \tau| \geq \Delta \tau_1$ eliminates correlation with the authentic
signal.

\section{Measurement Models}
\label{sec:measurement-models}
This section develops models for the two observations that constitute the
interference detection statistic, namely, the received power measurement and
the symmetric difference measurement.

\subsection{Received Power Measurement}
The total received power measured by a GNSS receiver in an RF band of
interest, denoted $P_k$ for measurement time $t_k$, is a simple and effective
indicator of interference \cite{ward1994aat,akos2012s}.  Even subtle spoofing
attacks with power advantage $\eta$ near unity can cause an increase in $P_k$
that is distinguishable from random variations, including so-called nulling
attacks that attempt to suppress the authentic signals
\cite{humphreysGNSShandbook}.  The catch is that routine events such as solar
radio bursts and the near passage of a so-called personal privacy device can
also cause $P_k$ to rise above nominal levels \cite{humphreysGNSShandbook}.
Thus, to prevent an anomalous-power monitor from issuing alarms intolerably
often, the alarm threshold may need to be raised to a point where the monitor
is no longer sensitive to subtle spoofing attacks.  This is why power
monitoring is best coupled with other forms of interference monitoring.

Suppose $M_s + 1$ authentic signals modeled as (\ref{eq:rA}) are received,
each with its associated interference signal, modeled as (\ref{eq:si}).  Let
$r_{{\rm A}i}(t)$ and $r_{{\rm I}i}(t)$ represent the $i$th authentic and
interference signals.  The combined signal exiting the RF front end is then
\begin{equation}
  \label{eq:totalReceivedPower}
  r_{\rm C}(t) = \sum_{i=0}^{M_s} \left[ r_{{\rm A}i}(t) + r_{{\rm I}i}(t)
  \right] + N(t)
\end{equation}
where $N(t)$ is the same thermal noise component as in (\ref{eq:rN_model}).  

\begin{figure}
  \centering
  \includegraphics[width=0.48\textwidth]{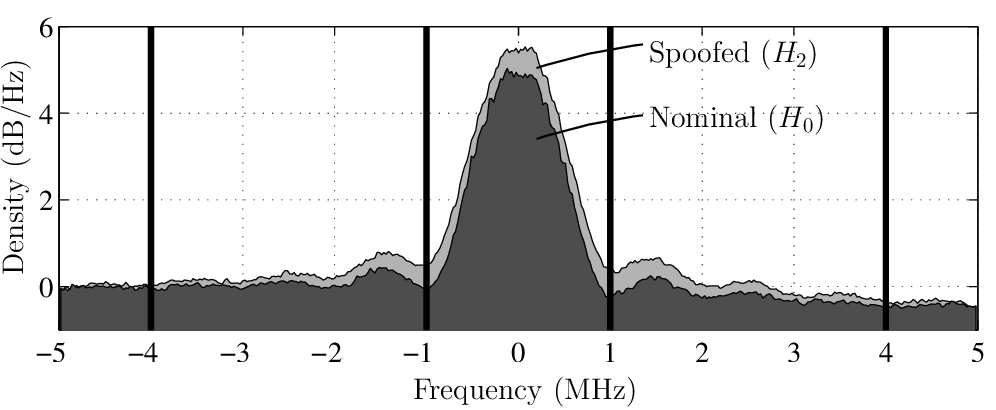}
  \caption{Normalized power spectral density about the GPS L1 C/A center
    frequency (1575.42~MHz) for a static receiver platform during nominal
    conditions in a clean RF environment (dark gray), and during a spoofing
    attack (light gray).  The two spectra are superimposed so that, of the
    spoofing attack spectrum, only the excess power beyond nominal is visible.
    The inner two vertical lines represent the 2-MHz bandwidth about L1, the
    outer two the 8-MHz bandwidth.  The spoofing attack spectrum applies at
    $t_k = 130$ seconds into scenario 7 of the Texas Spoofing Test Battery
    (TEXBAT {\tt tb7}), described in \cite{texbat_ds7_ds8}. }
  \label{fig:spectrum}
\end{figure}

\begin{figure}
  \centering
  \includegraphics[width=0.48\textwidth]{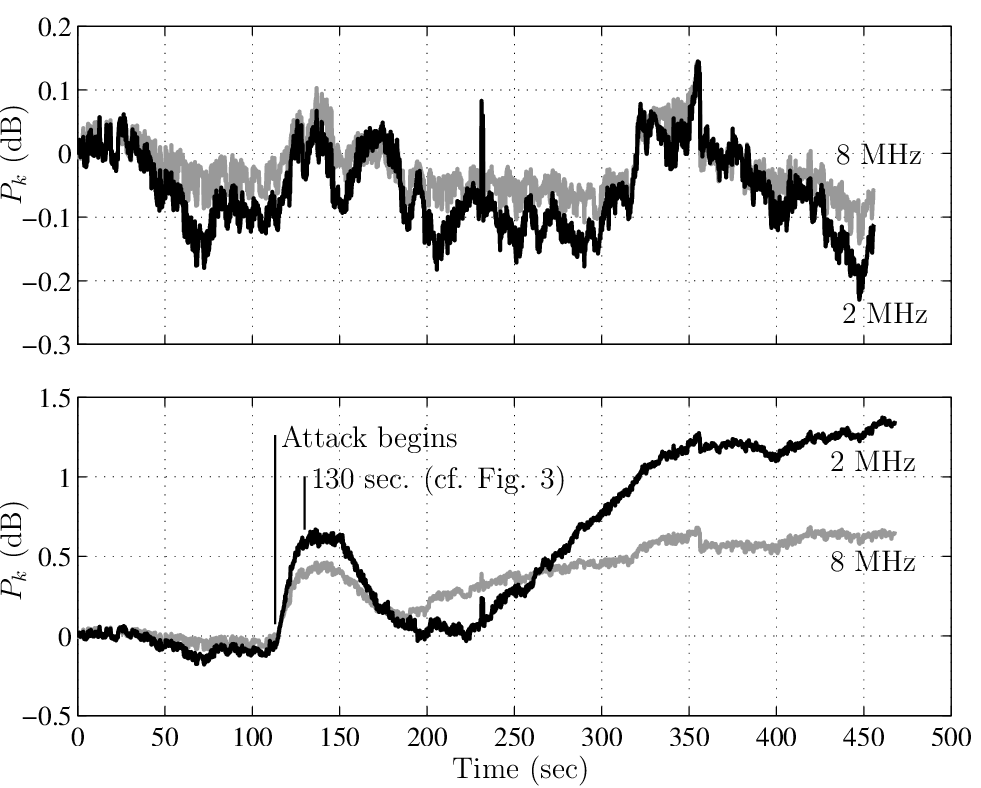}
  \caption{Time histories of measured received power $P_k$ for the nominal
    scenario (top) and spoofing scenario (bottom) whose spectra are shown in
    Fig. \ref{fig:spectrum}.  Black traces correspond to the 2-MHz bandwith
    and gray traces to the 8-MHz bandwidth centered at L1. The averaging
    interval for each measurement is 200 ms.  $P_k$ has been normalized to 0
    dB at $t_k = 0$.  The 130-second mark noted in the bottom panel is the
    instant at which the spoofing attack spectrum in Fig. \ref{fig:spectrum}
    applies.  At this point in the attack, $\eta_i \approx 4$,
    $\Delta \theta_i \approx \pi$, and $\Delta \tau_i \approx 0$ for
    $i = 0,...,M_s$.}
  \label{fig:power}
\end{figure}

Receivers with sufficient dynamic range in the discrete samples produced from
$r_{\rm C}(t)$ do not require an AGC and so can compute received power
directly by averaging the squared modulus of these samples.  Let $W_{P}$ be
the bandwidth over which $P_k$ is to be measured.  If $W_P$ is narrower than
the receiver front-end's native bandwidth, then $r_{\rm C}(t)$ must be
filtered to isolate the desired spectral interval.  Fig. \ref{fig:spectrum}
shows 2- and 8-MHz bands for $P_k$, with Fig. \ref{fig:power} showing
corresponding power time histories.  In a receiver whose front end is equipped
with an AGC, $P_k$ is measured indirectly through the AGC setpoint
\cite{akos2012s}.  In this case $W_{P}$ is equivalent to the front-end
noise-equivalent bandwith.

Let $\tilde{r}_{\rm C}(t)$ represent the (optionally) filtered version of
$r_{\rm C}(t)$ from which $P_k$ is to be measured.  $P_k$ is calculated as the
average power in $\tilde{r}_{\rm C}(t)$ over the interval $t_{k-1}$ to $t_k$:
\begin{align}
  P_k ~\mbox{(dBW)} \equiv 10 \log_{10} \left(\frac{1}{T} \int_{t_{k-1}}^{t_k}
  |\tilde{r}_{\rm C}(t)|^2 dt \right)
\end{align}

Let $p_{P_k}(x | \theta)$ be the conditional density of $P_k$. Assuming $W_P$
is wide enough that $\tilde{r}_{\rm C}(t)$ retains the significant power in
$r_{{\rm A}i}(t)$ and $r_{{\rm I}i}(t)$, then $p_{P_k}(x | \theta)$ can be
modeled as Gaussian, i.e.,
$p_{P_k}(x | \theta) = \mathcal{N}(x; \bar{P}_k, \sigma_P^2)$, where
$\bar{P}_k$ and $\sigma_P$ are expressed in dBW. The mean, $\bar{P}_k$,
revealed by carrying out the multiplications inherent in
$|\tilde{r}_{\rm C}(t)|^2 = \tilde{r}_{\rm C}(t)\tilde{r}_{\rm C}^*(t)$, and
by noting that $N(t)$ is zero-mean and independent of $r_{{\rm A}i}(t)$ and
$r_{{\rm I}i}(t)$, and that, due orthogonality of spreading codes,
$E[r_{{\rm A}i}(t)r_{{\rm A}i}^*(t)] = E[r_{{\rm I}i}(t)r_{{\rm I}l}^*(t)] =
0$ for all $i \neq l$, is $\bar{P}_k = 10 \log_{10}(\bar{P}_{\rm L})$, where
\begin{align}
\label{eq:Pk_model}
\bar{P}_{\rm L} = & \sum_{i = 0}^{M_s} 
   \left[(1+\eta_i)P_{{\rm A}i} 
   + 2\sqrt{\eta_i}P_{{\rm A}i}\cos(\Delta
    \theta_i)R(\Delta \tau_i)\right] \\ & + N_0 W_{P} \nonumber
\end{align}
and the subscript $i$ associates the corresponding quantity with the $i$th
signal.  The first term in the summation is the non-coherent sum of power in
$r_{{\rm A}i}$ and $r_{{\rm I}i}$; the second term is the power contributed
by coherent interaction between $r_{{\rm A}i}$ and $r_{{\rm I}i}$.  If
$\eta_i = 0$, or if the interference is orthogonal to the authentic signal
(e.g., $\Delta \theta_i = \pm \pi/2$), or if $\Delta \tau_i$ is large enough
that $R(\Delta \tau_i) = 0$, then this second term vanishes. 

The deviation $\sigma_P$ about $\bar{P}_k$ accounts for unpredictable but
natural variations in $P_k$ such as measurement error due to the finite
duration over which $P_k$ is estimated.  From the top plot of
Fig. \ref{fig:power}, one can see that for the receiver under test, $\sigma_P$
is at approximately 0.1 dB and 0.05 dB, respectively, for 2- and 8-MHz
bandwidths.  For urban RF environments, in which significant low-level
interference is present in the GNSS bands, $\sigma_P$ can be as high as 0.5 dB
in a 2-MHz bandwidth.  This paper's detection test assumes $\sigma_P = 0.4$
dB.

For modeling completeness, the multi-access interference $M(t)$ in
(\ref{eq:rN_model}) can be interpreted in terms of the model for
$r_{\rm C}(t)$ in (\ref{eq:totalReceivedPower}).  $M(t)$ models the effect of
multi-access interference for a particular desired signal, $r_{\rm A}(t)$.
Invoking the so-called thermal-noise approximation
\cite{humphreysGNSShandbook}, $M(t)$ is assumed to be spectrally flat with
density $M_0$, whose variable value is a function of the spreading code's chip
interval, $\tau_c$, and of the average power $\bar{P}_{\rm M}$ and number
$M_s$ of multi-access signals.  For GNSS signals with a
$\sinc^2(f\tau_c)$-shaped power spectrum, which applies for all GPS signals
except the new military M-code signals, $M_0$ is given by
\cite{humphreysGNSShandbook}
\begin{align}
\label{eq:M0_model}
  M_0 = (2/3)M_s \bar{P}_{\rm M} \tau_c
\end{align}
Three example cases for $\bar{P}_{\rm M}$ are given below.  In each case,
$i = 0$ indicates the desired authentic signal and its interferer, whereas
$i = 1,...,M_s$ indicate multi-access signals and their interferers.
\begin{description}[leftmargin=0.3cm,style=sameline,font=\normalfont]
\item[$\bar{P}_{\rm M} = P_{\rm A}$:] In this case, $P_{{\rm A}i} = P_{\rm A}$
  for $i = 0,...,M_s$ and $\eta_i = 0$ for $i = 1, ..., M_s$.  In other words,
  $P_{\rm A}$ is assumed to be typical of the received power from all GNSS
  signals and no other interference besides $r_{{\rm I}0}$ is assumed.  This
  case applies for $H_0$, for $H_1$ when only a single signal, $r_{{\rm A}0}$,
  experiences significant multipath, and for $H_2$ when the spoofing attack
  targets only $r_{{\rm A}0}$, as opposed to an attack against all
  $r_{{\rm A}i}$.
\item[$\bar{P}_{\rm M} = (1+\eta)P_{\rm A}$:] In this case,
  $P_{{\rm A}i} = P_{\rm A}$ and $\eta_i = \eta$ for $i = 0,...,M_s$.  In
  other words, each of the $M_s +1$ received signals is modeled as having
  power $P_{\rm A}$, and each is assumed to be subject to a non-coherent
  interferer with power $\eta P_{\rm A}$.  This case applies for the jamming
  hypothesis $H_3$, as follows: For all $i$, $|\Delta \tau_i|$ is assumed to
  be large enough that $R(\Delta \tau_i) = 0$, causing the second term of the
  summation in (\ref{eq:Pk_model}) to vanish.  Jamming power manifests in the
  first term of the summation as an additional $\eta_i P_{{\rm A}i} = \eta P_{\rm A}$.

  This case also applies for $H_2$ when the spoofing-to-authentic phase
  difference $\Delta \theta_i$ for $i = 0,...,M_s$ can be modeled as a random
  variable uniformly distributed on $[0,2 \pi)$, in which case the second term
  of the summation in (\ref{eq:Pk_model}) vanishes for all $i$.
\item[$\bar{P}_{\rm M} = (1+\eta)P_{\rm A} + 2\sqrt{\eta}P_{\rm A}\cos(\Delta
  \theta)R(\Delta \tau)$:] This case applies for $H_2$ when the spoofing
  attack targets not only $r_{{\rm A}0}$ but also the $M_s$ multi-access
  signals such that $\eta_i P_{{\rm A}i} = \eta P_{\rm A}$,
  $\Delta \theta_i = \Delta \theta$ and $\Delta \tau_i = \Delta \tau$ for
  $i=0,...,M_s$.  
\end{description}

\subsection{Symmetric Difference Measurement}
\begin{figure}
  \centering
  \includegraphics[width=0.5\textwidth, trim={0.9cm 0.3cm 0.7cm 0.7cm},clip]
  {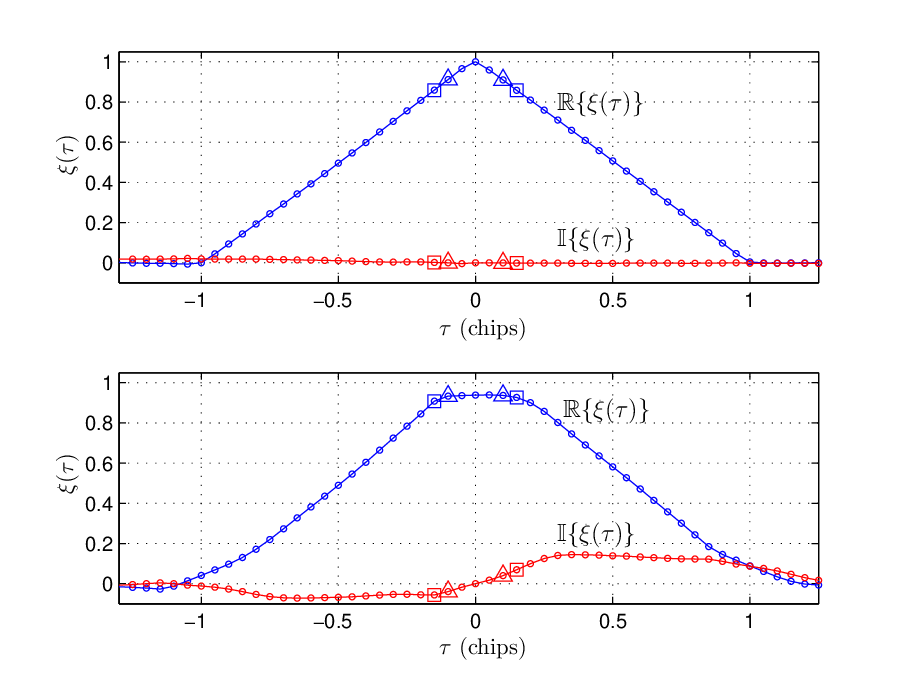}
  \caption{In-phase $\mathds{R}\{\xi_k(\tau)\}$ and quadrature
    $\mathds{I}\{\xi_k(\tau)\}$ components of the empirical correlation
    function $\xi_k(\tau)$ for a clean scenario (top) and for a spoofing
    scenario (bottom).  The early and late taps used for tracking
    [$\xi_k(-\tau_{\rm DLL})$ and $\xi_k(\tau_{\rm DLL})$; marked with
    $\square$] and for the symmetric difference measurement
    [$\xi_k(-\tau_{d})$ and $\xi_k(\tau_{d})$; marked with $\triangle$] are
    not necessarily the same. The nominal case (top) exhibits symmetry about
    $\tau = 0$ in both components, whereas the spoofed case (bottom) shows
    distortion and strong asymmetry in both components.  Samples of the signal
    $r(t)$ have been scaled such that $|\xi_k(0)| = 1$ for the nominal case.}
  \label{fig:autocorr}
\end{figure}

A variety of signal quality monitoring (SQM) metrics have been applied to
detect distortions in $\xi_k(\tau)$ caused by anomalous GNSS signals,
including spoofing signals \cite{r_phelts2001sqm, m_irsigler2005mco,
  gunawardena2009anomalous, cavaleri2010dscc,
  ledvina2010cgr,o_mubarak2010elp,gamba2016hypothesis}.  Among these, the
so-called symmetric difference is particularly attractive for multipath and
spoofing detection because it is simple to implement, is insensitive to the
particular shape of the correlation function (provided the function is
symmetric about $\tau = 0$, as is the case for all current and proposed GNSS
signals), and is sensitive to the correlation function distortion introduced
by spoofing whenever the authentic and spoofing signals are approximately
matched in power (i.e., $0.1 < \eta < 10$).  Fig. \ref{fig:autocorr}
illustrates $\xi_k(\tau)$ under nominal conditions (top) and a spoofing attack
(bottom).

Let $\tau_d$ be the offset of the early and late symmetric difference taps
from $\tau = 0$, and let $\sigma_{{\rm N}0}$ be the value of $\sigma_{\rm N}$
in the interference-free case, for which we assume $M_0$ is given by
(\ref{eq:M0_model}) with $\bar{P}_{\rm M} = P_{\rm A}$.  Then at measurement
time $t_k$ the symmetric difference is calculated as the modulus of the
complex difference between the early and late symmetric difference taps,
scaled by $1/\sigma_{{\rm N}0}$:
\begin{align}
  D_k(\tau_d) \triangleq \frac{| \xi_k(-\tau_d) - \xi_k(\tau_d)|}{{\sigma}_{{\rm N}0}}
  \label{eq:sd}
\end{align}

As with $P_k$, a statistical model for $D_k(\tau_d)$ is necessary to develop a
Bayesian detection framework. Let $p_{D_k}(x | \theta)$ denote the conditional
density of $D_k$.  An analytical model for $p_{D_k}(x | \theta)$ is easily
derived when there is no coherent interaction between $r_{\rm A}$ and
$r_{\rm I}$, that is, when the second term of the summation in
(\ref{eq:Pk_model}) vanishes, which occurs in the interference-free case
($\eta = 0$) or in the case of jamming, for which $R(\Delta \tau) = 0$.  Let
$P_{H_0}$ denote the received power when $\eta = 0$, which, as evident in the
top panel of Fig. \ref{fig:power}, is constant to within a few tenths of a dB.
Assume that, for any $r(t)$ and any $k \in \{1,2,...\}$, the AGC adjusts
$\beta_k$ such that the average power in $\beta_k r(t)$ over $t_{k-1}$ to
$t_k$ is $P_{H_0}$.  Then, if code tracking is accurate
($\Delta \tau_{{\rm A}k} \approx 0$), as would be true for up to moderate
jamming, or if the amplitude of $\xi_A$ is small with respect to
$\sigma_{\rm N}$, as would be true for strong jamming, then the scaled complex
difference $[\xi_k(-\tau_d) - \xi_k(\tau_d)]/\sigma_{{\rm N}0}$ can be modeled
as a zero-mean complex Gaussian random variable with variance
\begin{align}
  \label{eq:sigmad}
  \sigma_d^2 = 8 \tau_d  \left(\frac{ \beta_k\sigma_{\rm N}}{\sigma_{{\rm N}0}}\right)^2
\end{align}
It follows that $p_{D_k}(x | \theta)$ in this case is Rayleigh with scale
parameter $\frac{\sigma_d}{\sqrt{2}}$:
\begin{align}
  \label{eq:rayleigh}
  p_{D_k}(x | \theta) = \frac{2x}{\sigma_d^2}\exp{\left(\frac{-x^2}{
  \sigma_d^2}\right)}, \quad x \geq 0
\end{align}
In the interference-free case, $\beta_k = 1$ and
$\sigma_{\rm N} = \sigma_{{\rm N}0}$, so $\sigma_d^2$ reduces to
$\sigma_d^2 = 8\tau_d$.  One can show that this simplified expression for
$\sigma_d^2$ is also approximately true for wideband jamming, for which the
AGC maintains $(\beta_k \sigma_{\rm N}/\sigma_{{\rm N}0})^2 \approx 1$.
However, the structured-signal jamming assumed in this paper is more potent
than wideband jamming: it increases $M_0$, and therefore $\sigma_{\rm N}$,
more than equivalently-powered wideband jamming \cite{humphreysGNSShandbook}.
As a result, $\sigma_d^2$ rises with increasing power of structured-signal
jamming, making it more difficult to distinguish jamming from spoofing.
Nevertheless, assuming structured-signal jamming, as opposed to the less
potent wideband or narrowband jamming, is consonant with this paper's focus on
worst-case attacks.

Note that a version of the symmetric difference normalized by the maximum
magnitude of $\xi_k(\tau)$ has also been proposed for SQM and spoofing
detection \cite{r_phelts2001sqm, m_irsigler2005mco, huang2016gnss}.  But the
definition of $D_k$ in (\ref{eq:sd}) is more convenient because its
distribution for $\eta = 0$ depends only on $\sigma_d$, as evident in
(\ref{eq:rayleigh}).

Modeling $p_{D_k}(x | \theta)$ as Rayleigh only holds for $\eta = 0$ or
$R(\Delta \tau) = 0$.  When neither of these conditions is true, the real and
imaginary components of $\xi_k(\tau)$ can manifest strong asymmetry, as shown
in the bottom panel of Fig.~\ref{fig:autocorr}.  In this case, the possible
interactions of $\xi_{{\rm I}k}(\tau)$, and $\xi_{{\rm A}k}(\tau)$ are so
complex that $p_{D_k}(x | \theta)$ cannot be modeled analytically.  It can,
however, be studied through Monte-Carlo simulation of (\ref{eq:af}) and
(\ref{eq:sd}) as a function of $\theta = [\eta, \Delta \tau, \Delta \theta]^T$
and $\tau_d$, together with some assumed recipe for determination of
$\hat{\tau}$, the receiver's estimate of $\tau_{\rm A}$.  For this paper, we
have assumed that
\begin{equation}
  \label{eq:codePhaseEstimate}
 \hat{\tau} = \arg \max_\tau |\xi_k(\tau)| 
\end{equation}
which holds for early-minus-late DLL tracking (both coherent and non-coherent)
in the limit as $\tau_{\rm DLL} \rightarrow 0$, provided the tracking points
have not settled on a local maximum different from the global maximum.  Note
that to minimize the number of correlation taps, one can choose
$\tau_{\rm DLL} = \tau_d$, where $\tau_{\rm DLL}$ is the offset from
$\tau = 0$ of the DLL's early and late taps. But, depending on the shape of
$\xi_k(\tau)$ and the assumed distribution of $\tau_{\rm I}$, unequal
$\tau_{\rm DLL}$ and $\tau_d$ may improve tracking and detection performance.

In simulation, one can apply whatever algorithm for producing $\hat{\tau}$
best approximates the operation of the receiver, or class of receivers, whose
reaction to interference one wishes to simulate.  For example, a
multipath-mitigating receiver may estimate $\tau_{\rm A}$ by (1) assuming the
presence of one or more multipath components, (2) estimating the parameters
$\eta, \tau_{\rm I}$ and $\theta_{\rm I}$ for each component, and (3)
subtracting a model of each component from $\xi_k(\tau)$ \cite{weill2002mm}.

\subsection{Combined Measurement Vector}
The symmetric difference for tracking channel $i\in\{1, 2, ..., N\}$ at time
$t_k$, denoted $D_k^i(\tau_d)$, can be combined with the received power
measurement to form a channel-specific vector of observables at $t_k$:
\begin{align*}
  {z}_k^{i} = [D_k^{i}(\tau_d), P_k]^T
\end{align*}
$D_k^{i}(\tau_d)$ and $P_k$ are modeled as independent so that
\[ p(z_k^{i} | \theta) = p_{D_k}(z_{1k}^{i} | \theta) p_{P_k}(z_{2k}^{i} |
  \theta) \] where $z_{jk}^{i}$ is the $j$th element of $z_k^{i}$.

Obviously, when multiple GNSS signals are simultaneously affected by
interference, detection performance improves by considering the multi-channel
observation vector
\begin{align}
  \label{eq:combined}
  {z}_k^{(1:N)} = [D_k^1(\tau_d), D_k^2(\tau_d), \ldots, D_k^{N}(\tau_d),
  P_k]^T
\end{align}
For clarity of presentation and analysis, the channel-specific vector
${z}_k^i$ has been taken as the observation vector for this paper's Bayesian
detection strategy, with the superscript $i$ dropped for notational
simplicity.  Extension of this paper's methods to the multi-channel case
follows straightforwardly from the principles of the per-channel test along
with a model for correlation between the $D_k^i(\tau_d)$.

\section{The Power-Distortion Tradeoff Under Spoofing}
\label{sec:power-dist-trad}
The foregoing measurement models allow one to appreciate the power-distortion
tradeoff that a spoofer faces when mounting an attack.  In a typical attack,
an admixture of spoofing and authentic signals is incident on the receiver's
antenna.  If the spoofing and authentic signals are approximately matched in
power (i.e., $0.1 < \eta < 10$), then the correlation function will be
significantly distorted, as shown in the lower panel of
Fig.~\ref{fig:autocorr}.  The spoofer has several options for reducing this
telltale distortion, but the alternatives are either intrinsically difficult,
lead to anomalously high received power, or preclude effective capture, as
described below.

\subsection{Nulling or Blocking}
\label{sec:nulling-or-blocking}
A spoofer can minimize the hallmark distortions of an attack by generating an
antipodal, or nulling, signal or by preventing reception of the authentic
signal (e.g., by emplacing an obstruction). The form given for $r_{\rm I}(t)$
in (\ref{eq:si}) accommodates a nulling attack in which $r_{\rm I}(t)$ is made
antipodal to $r_{\rm A}(t)$ via the following settings: $\eta = 1$,
$\tau_{\rm I} = \tau_{\rm A}$, and $\theta_{\rm I} = \theta_{\rm A} + \pi$.
This attack annihilates $r_{\rm A}(t)$, producing an effect similar to jamming
but with much less received power.  But the form in (\ref{eq:si}) does not
accommodate the nulling-and-replacement attack described in
\cite{psiaki2016gnssSpoofing} and \cite{humphreysGNSShandbook}, whereby, after
nulling $r_{\rm A}(t)$, the attacker supplants it with a separate spoofing
signal.  Detection of a perfectly-executed nulling-and-replacement attack is,
in fact, not possible with the technique developed in this paper.  Such an
attack is, however, extremely difficult to carry out in practice, as it
requires centimeter-level knowledge of, and a highly accurate fading model
for, the attacker-to-receiver signal path.

Similarly, blocking reception of $r_{\rm A}(t)$ by physical obstruction is
difficult in cases where the receiver antenna is not physically accessible to
the attacker.  The technique developed in this paper makes the assumption that
a nulling-and-replacement attack is impractically difficult and that physical
access to the receiving antenna is controlled to prevent signal blockage.  It
must be recognized, however, that in many cases of practical interest (e.g.,
fishing vessel monitoring), the receiver and antenna are fully under the
control of potential attackers.

\subsection{Overpowering}
An alternative approach to eliminating telltale distortion is an overpowered
spoofing attack in which $\eta$ is set so high that the receiver's AGC
squelches the relatively weak authentic signals below the noise floor, thereby
eliminating interaction between $r_{\rm I}$ and $r_{\rm A}$.  In other words,
as $\eta$ increases, $\beta_k$ decreases to maintain a constant power in
$\beta_k r(t)$, with the result that $\beta_k r_{\rm A}(t)$ becomes negligible
compared to $\beta_k r_{\rm I}(t)$.  If, however, the receiver raises an alarm
when the received power $P_k$ exceeds some threshold, then covert spoofing is
strictly limited to $\eta < \eta_{\rm max}$ for some $\eta_{\rm max}$.

\subsection{Underpowering}
The spoofer can also minimize correlation function distortion by selecting a
small $\eta$. However, reliable capture of the receiver's tracking loops
requires $\eta > \eta_{\rm min} \approx 0.4~$dB \cite{d_Shepard2011asa}. Thus,
for spoofing to be both reliable and covert, $\eta$ is lower bounded by
$\eta_{\rm min}$ and upper bounded by $\eta_{\rm max}$.

This paper's approach to spoofing detection can be stated as follows: If the
defending receiver is designed such that $\eta_{\rm max}$ is sufficiently low
despite maintaining a tolerable rate of false alarm in the received power
monitor, then a carry-off spoofing attack that respects this bound yet
successfully captures the receiver's tracking loops will unavoidably and
detectably distort the correlation function $\xi_k(\tau)$.  This distortion,
evident in $D_k(\tau_d)$, can be distinguished from distortion due to
multipath by its greater magnitude and by a concomitant increase in $P_k$.

\section{Decision Rule}
Finding the decision rule $\delta({z}_k)$ from (\ref{eq:bayesOrAverageRisk})
that minimizes the Bayes risk $r(\delta)$ is conceptually straightforward.
One can express $r(\delta)$ as
\begin{align}
\label{eq:bayesRisk2}
  r(\delta) = E\{C[\delta({Z}_k),{\Theta}]\} = 
  E\{E\{C[\delta({Z}_k),{\Theta}]|{Z}_k\}\}
\end{align}
where the second equality follows from the rule of iterated expectations.  We
note from (\ref{eq:bayesRisk2}) that, whatever the distribution of ${Z}_k$,
$r(\delta)$ is minimized when, for each ${z}_k \in \Gamma$, $\delta$ is
chosen to minimize the posterior cost
$E\{C[\delta({Z}_k),{\Theta}]|{Z}_k = {z}_k\}$.  Thus, the Bayes-optimal
rule is given by
\begin{equation}
  \label{eq:bayesRuleOptimal}
  \delta_{\rm B}({z}_k) = \arg \min_{i\in \mathcal{I}} E\{C[i,{\Theta}]|{Z}_k = {z}_k\}
\end{equation}
which, because the parameter sets $\Lambda_i$ are disjoint, can be written
\begin{equation}
  \label{eq:bayesRuleExpanded}
  \delta_{\rm B}({z}_k) = \arg \min_{i\in \mathcal{I}} \sum_{j=0}^3 \int_{\Lambda_j}
  C[i,{\theta}]p(\theta | z_k) d\theta
\end{equation}
Reversing the conditioning of $p(\theta | z_k)$ using Bayes's formula, and
recognizing that $w(\theta) = \sum_{j = 0}^3 w_j(\theta) \pi_j$, we obtain
\begin{equation}
  \label{eq:bayesRuleExpanded2}
  \delta_{\rm B}({z}_k) = \arg \min_{i\in \mathcal{I}} \sum_{j=0}^3 \pi_j \int_{\Lambda_j}
  C[i,{\theta}] p(z_k | \theta) w_j(\theta)  d\theta
\end{equation}

\begin{figure}[t]
  \centering
  \includegraphics[width=0.5\textwidth,trim={0.5cm 0 0.2cm 0.7cm},clip]
  {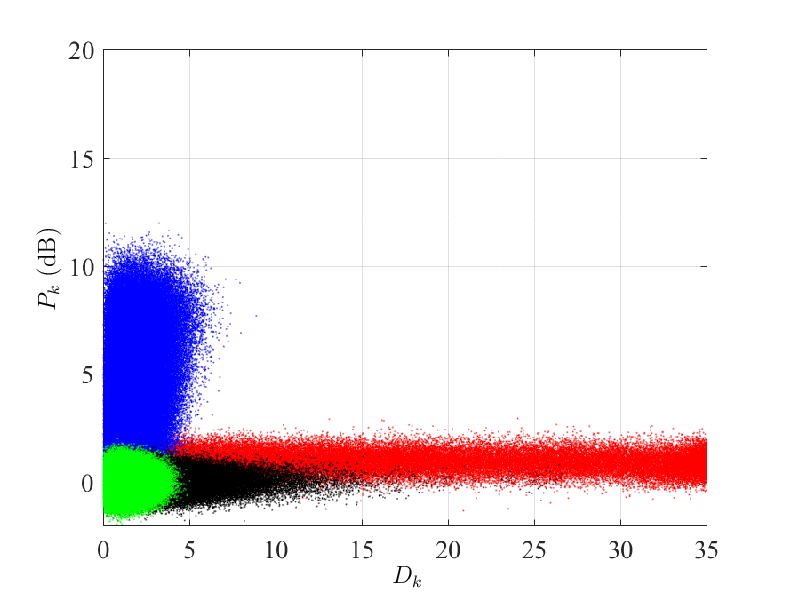} 
  \caption{Simulated observations ${z}_k = [D_k(\tau_d),P_k]^T$ for no
    interference (green), multipath (black), spoofing (red), and jamming
    (blue) based on the Monte-Carlo simulation technique described in the
    text, with
    $N_{\rm P} = 10^5, N_{\rm M} = 20, \pi_0 = 0.6, \pi_1 = 0.2, \pi_2 =
    0.05$, $\pi_3 = 0.15$, $\sigma_P = 0.4$ dB, $P_{\rm A} = -156$ dBW,
    $N_0 = -204$ dBW/Hz, $M_s = 7$, $\tau_d = 0.15$ chips, $T = 100$ ms,
    $W_{P} = 2$ MHz, and with $\bar{P}_{\rm M} = P_{\rm A}$ for $H_0$ and
    $H_1$,
    $\bar{P}_{\rm M} = (1+\eta)P_{\rm A} + 2\sqrt{\eta}P_{\rm A}\cos(\Delta
    \theta)R(\Delta \tau)$ for $H_2$, and
    $\bar{P}_{\rm M} = (1+\eta)P_{\rm A}$ for $H_3$.  Note that the spoofing
    samples are concentrated along a low-power band because the distribution
    $w_2(\theta)$ assumes stealthy spoofing that attempts to masquerade as
    multipath.}
  \label{fig:sim}
\end{figure}
\begin{figure}[t]
  \centering
  \includegraphics[width=0.48\textwidth,trim={0.6cm 0 0.7cm 0.3cm},clip]
  {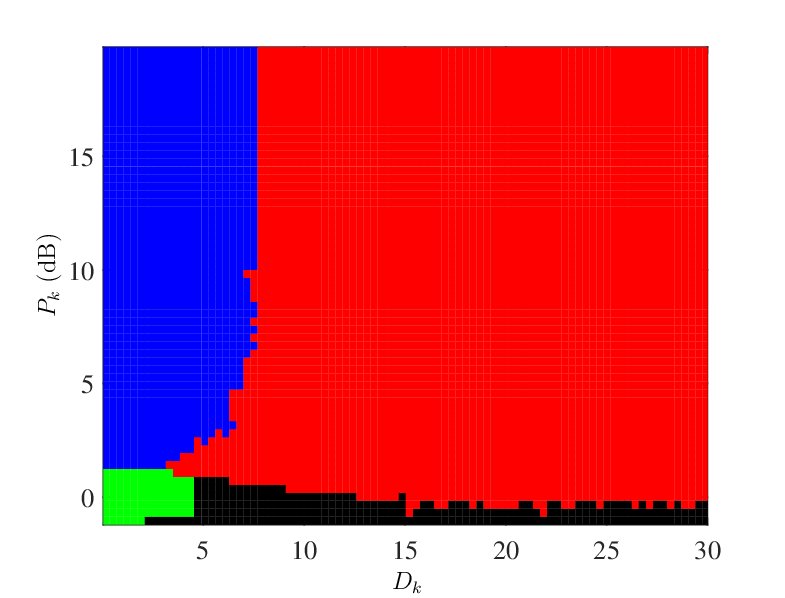}
  \caption{Optimum decision regions for the $\theta$-dependent cost
    $C[i,\theta]$: $\Gamma_0$ (no interference, green), $\Gamma_1$ (multipath,
    black), $\Gamma_2$ (spoofing, red), and $\Gamma_3$ (jamming, blue).}
  \label{fig:regions}
\end{figure}

The foregoing sections defined the parameter sets $\Lambda_i$ and presented
analytical models for the conditional densities $w_i(\theta)$,
$i\in \mathcal{I}$.  If an analytical model for $p(z_k | \theta)$ were also
available, then (\ref{eq:bayesRuleExpanded2}) could be solved by numerical
integration.  Unfortunately, finding an analytical model for $p(z_k | \theta)$
only appears possible for the special cases of $\eta = 0$ or
$R(\Delta \tau) = 0$.  On the other hand, for a given $\theta$, it is easy to
simulate samples drawn from $p(z_k | \theta)$ by taking the nonlinear and
stochastic models described in Sections \ref{sec:sign-interf-model} and
\ref{sec:measurement-models} as recipes for a sample simulator.  In view of
this, a Monte-Carlo technique can be applied to find $\delta_{\rm B}(z_k)$, as
follows.
\begin{enumerate}
\item For each $i\in\mathcal{I}$, $N_i$ parameter vectors are simulated
  according to the distribution $w_i(\theta)$, where $N_i/N_{\rm P} \approx \pi_i$
  and $N_{\rm P} = \sum_{i \in \mathcal{I}} N_i$.  The $l$th simulated vector drawn from
  $w_i(\theta)$ is denoted ${\theta}_{l i}$.
\item For each ${\theta}_{l i}$, $N_{\rm M}$ simulated measurements ${z}_k$ are
  generated.  Dropping the time index $k$ from $z_k$ for notational clarity,
  the $m$th simulated measurement, given ${\theta}_{l i}$, is written
  ${z}_{mli}$.  The total number of measurements simulated is $N_{\rm P} N_{\rm M}$.
  Fig. \ref{fig:sim} shows an example realization of this sample generation
  process over a relevant range for $D_k$ and  $P_k$.
\item The two-dimensional observation space $\Gamma$ is divided into a large
  number of small rectangular cells of uniform size.  Each cell is assumed to
  belong to a single decision region such that all observation samples falling
  within a cell belonging to $\Gamma_i$ are assigned to hypothesis
  $H_i,~i \in \mathcal{I}$. An initial partition of $\Gamma$ is created by
  assigning each cell to the hypothesis $H_i$ having the largest number of
  samples ${z}_{mli}$ within the cell. Boundaries are adjusted such that each
  decision region $\Gamma_i$ is simply connected (no islands), a condition
  that can reasonably be assumed from the fact that $C[i,\theta]$,
  $p(z_k|\theta)$, and $w_i(\theta)$ are smooth in $\theta$.
\item The Bayes risk for the current partition is calculated as
  \begin{equation}
    \label{eq:bayesOperational}
  r(\delta) = \frac{1}{N_{\rm P} N_{\rm M}} \sum_{i=0}^3 \sum_{l=1}^{N_i}
    \sum_{m=1}^{N_{\rm M}} C[\delta({z}_{mli}), {\theta}_{l i} ] 
  \end{equation}
  A new decision region assignment is considered for each cell lying along the
  boundary between decision regions.  The new assignment is retained whenever
  it reduces $r(\delta)$, provided the resulting decision regions remain
  simply connected.  The process is repeated until no boundary cells warrant
  re-assigning, at which point the cell assignments constitute the final
  decision regions $\Gamma_i, i \in \mathcal{I}$.  Fig. \ref{fig:regions}
  shows an example of the decision regions created by this process.
\end{enumerate}

Two different types of cost function $C[i,{\theta}]$ are considered, as
follows.
\subsection{Uniform Cost within each $\Lambda_j$}
\label{sec:cost}
When $C[i,\theta]$ is uniform across all ${\theta} \in \Lambda_j$, we write
$C_{ij}$, interpreted as the cost of choosing $H_i$ when $H_j$ is true.  The
simplest such cost evenly penalizes misclassification: if $i = j$, then
$C_{ij} = 0$; otherwise, $C_{ij} = 1$.  But in the context of navigation
security, not all types of misclassification are equally costly.  We propose
the following cost assignment:
\begin{description}[leftmargin=0.3cm, style=sameline, font = \normalfont]
\item[$C_{ii} = 0,~i \in \mathcal{I}$:]  Correct decision
\item[$C_{01} = 0.2$:] Low cost: Some multipath-induced code- and carrier-
  phase error might have been mitigated via receiver's multipath mitigation
  routines had multipath been detected, but otherwise harmless.
\item[$C_{02} = C_{12} = 1$:] Highest cost: Spoofing goes undetected;
  receiver may report hazardously misleading information. Multipath mitigation
  applied by a receiver deciding $H_1$ cannot be assumed to reduce this cost.
\item[$C_{03} = C_{13} = 0.9$:] High cost: Jamming goes undetected; receiver
  may report hazardously misleading information. Not so costly as $C_{02}$
  because a jammer has less control over receiver output than a spoofer.
\item[$C_{10} = 0.1$:] Low cost: Receiver may waste computational resources
  trying to mitigate phantom multipath, and runs a chance of slightly biasing
  code and carrier measurements in the process, but otherwise harmless.
\item[$C_{20} = C_{21} = C_{30} = C_{31} = 0.4$:] Moderate cost:
  A false alarm for jamming or spoofing is raised, breaking navigation
  solution continuity.
\item[$C_{32} = C_{23} = 0.2$:] Low cost: Although spoofing is misclassified
  as jamming or vice-versa, no hazardously misleading information is issued,
  as the receiver suppresses its navigation solution when it decides $H_2$ or
  $H_3$.
\end{description}

\subsection{$\theta$-Dependent Cost}
\label{sec:costTheta}
The harm to a GNSS receiver and its dependent systems or users may not be
uniform across all ${\theta} \in \Lambda_i$ for a given
$\Lambda_i \subset \Lambda,~ i\in \mathcal{I}$.  It is therefore worthwhile to
consider the full $\theta$-dependent cost $C[i,{\theta}]$.  Our approach
begins with the uniform cost assignment described above, substituting
$C[i,\theta], ~\theta \in \Lambda_j$ for certain elements $C_{ij}$, as
follows:
\begin{description}[leftmargin=0.3cm, style=sameline, font = \normalfont]
\item[{$C[0,\theta \in \Lambda_1] = \min[0.8,e_\tau(\theta)/0.3]$}:] Not all
  multipath is equally harmful: Cost is assumed to increase linearly with the
  magnitude of multipath-induced code-phase error
  $e_\tau(\theta) \equiv |\hat{\tau}(\theta) - \tau_A|$, expressed in chips of
  length $\tau_c$, up to a saturation value of $0.8$, where
  $\hat{\tau}(\theta)$ is given by (\ref{eq:codePhaseEstimate}) and
  $\min[x,y]$ is the minimum of $x$ and $y$.
\item[{$C[0,\theta \in \Lambda_2] = C[0,\theta \in \Lambda_1]$}] and
  $C[1,\theta \in \Lambda_2] = 0$ if
  $\Delta \tau < \tau_{\rm DLL}~ \mbox{and} ~\eta < -1$ dB; otherwise,
  $C[0,\theta \in \Lambda_2] = C_{02}$ and
  $C[1,\theta \in \Lambda_2] = C_{12}$, where $\tau_{\rm DLL}$ is the offset
  of the correlation taps used for tracking (marked with $\square$ in
  Fig. \ref{fig:autocorr}): Close-in, weak spoofing is no more harmful than
  multipath, and can be treated accordingly.
\item[{$C[0,\theta \in \Lambda_3] = C[1,\theta \in \Lambda_3] =$}]
  $\min[C_{03},\eta/10]$: The cost of jamming is assumed to increase linearly
  with $\eta$ (in dB) up to saturation at $C_{03}$.
\end{description}

\subsection{Prior Probabilities}
\label{sec:prior-probabilities}
The prior probabilities $\pi_i, ~i \in \mathcal{I}$, are of course
situation-dependent.  One may wish to increase $\pi_2$, for example, upon
entering an area where spoofers have historically been active.  Values of
$\pi_i, ~i \in \mathcal{I}$ were chosen as follows to represent a heightened
threat scenario.  Approximate relative values for $\pi_0, \pi_1$, and $\pi_3$
were first found by a manual epoch-by-epoch classification of data from a
mobile GPS receiver in an urban environment, which exhibited significant
multipath and mild jamming but no spoofing.  A low but significant prior
probability of spoofing was assumed, and the empirical $\pi_1$ and $\pi_3$
values were slightly inflated.  All values were then normalized such that
$\sum_{i\in \mathcal{I}} \pi_i = 1$.  The resulting values, assumed for the
remainder of the paper, were $\pi_0 = 0.6, \pi_1 = 0.2, \pi_2 = 0.05$ and
$\pi_3 = 0.15$.

\subsection{Application of the Decision Rule}
\label{sec:optim-decis-regi}
The PD detector's Bayes-optimal decision rule $\delta_{\rm B}$ for classifying
received GNSS signals based on observations ${z}_k = [D_k(\tau_d), P_k]^T$,
priors $\pi_i$, densities $w_i(\theta)$, $p(z_k|\theta)$, and the
$\theta$-dependent cost $C[i,\theta]$, all as described previously, and with
parameter values as indicated in the caption of Fig. \ref{fig:sim}, is
embodied in the colored regions shown in Fig. \ref{fig:regions}.

Application of the decision rule is straightforward: If the observation
${z}^j_k$, taken at time $t_k$ from tracking channel $j$, falls in decision
region $\Gamma_i$, it is assigned to hypothesis $H_i$.  Table
\ref{tab:misclassSim} shows classification statistics as evaluated by applying
the rule to an independent set of Monte-Carlo-generated observations like the
one shown in Fig. \ref{fig:sim}.  The table reveals that $\delta_{\rm B}$
tends to misclassify multipath as $H_0$, a consequence of the low cost
$C[0,\theta \in \Lambda_1]$, especially when multipath is benign, as is often
the case. Multipath is misclassified as spoofing less than $1.7\%$ of the time
even though the modeled $w_2(\theta)$ is unfavorable for distinguishing $H_2$
from $H_1$.

One can achieve a lower spoofing false alarm rate by considering a set of
channel-specific decisions over time
$\{\delta_{\rm B}({z}^j_k) | k \in \mathcal{K} \}$ for some set $\mathcal{K}$
of contiguous sample times.  This approach is effective because a spoofing
attack must proceed slowly enough to capture the receiver tracking loops,
which offers a window of several seconds over which
$\{\delta_{\rm B}({z}^j_k) | k \in \mathcal{K} \}$ will include many
declarations of $H_2$, whereas, under $H_1$, only a small number of such
declarations would arise.  One may also lower the spoofing false alarm rate by
considering multi-channel decisions
$\{\delta_{\rm B}({z}^j_k) | j = 1, 2, ..., N \}$, or the full combined
observation in (\ref{eq:combined}).  When applying multi-channel tests, one
must bear in mind that a spoofer may not attack all channels simultaneously.
However, any spoofer wishing to evade simple RAIM-type alarms must produce a
self-consistent signal ensemble \cite{khanafseh2014gps}, which requires
spoofing more than $N - N_m$ signals, where $N$ is the total number of
independent signals being tracked, and $N_m$ (typically 4) is the minimum
number required for a position and time solution.

\begin{table}[h]
  \centering
  \caption{Simulation-evaluated classification matrix for the decision regions
    in Fig. \ref{fig:regions}.  The table's $(i,j)$th element is the relative
    frequency with which the detector chose $i$ when $j$ was the true scenario.}
  \begin{tabular}[c]{ccccc}
    \toprule
     Decision & \multicolumn{4}{c}{True Scenario} \\
    \cmidrule(r){1-1} \cmidrule(r){2-5}
        & $H_0$ & $H_1$ & $H_2$ & $H_3$ \\  \midrule     
    $H_0$ & 0.9942 & 0.8809 & 0.06244 & 0.0184 \\
    $H_1$ & 0.0005 & 0.0987 & 0.0234 & 0 \\
    $H_2$ & 0.0001 & 0.0162 & 0.8698 & 0.0017 \\
    $H_3$ & 0.0039 & 0.0028 & 0.0442 & 0.9799 \\
    \bottomrule
  \end{tabular}
\label{tab:misclassSim}
\end{table}

\section{Experimental Data}
\label{sec:experimental-data}
An independent evaluation of the PD detector was carried out against 27
empirical GNSS data recordings, including 6 recordings of various spoofing
scenarios, 14 multipath-dense scenarios, 4 jamming scenarios of different
power levels, and 3 scenarios exhibiting negligible interference beyond
thermal and multi-access noise.  Table~\ref{dataDescription} provides a
summary description of each recording.

\subsection{TEXBAT}
The Texas Spoofing Test Battery (TEXBAT), version 1.1, from which {\tt cd0}
and {\tt tb2}--{\tt tb7} are drawn, is a public set of high-fidelity digital
recordings of spoofing attacks against civil GPS L1 C/A signals
\cite{humphreys2012_TEST_Battery,rnlTexbatSite,texbat_ds7_ds8}. Both static
and dynamic scenarios are provided along with their corresponding un-spoofed
recording.  Each 16-bit quantized recording is centered at 1575.42~MHz with a
bandwidth of 20~MHz and a complex sampling rate of 25~Msps.  Each spoofing
scenario makes use of the most advanced civil GPS spoofer publicly disclosed
\cite{t_humphreys_gcs08,t_humphreys2014stb}.  In the laboratory, the spoofer
can precisely control $\eta, \Delta \tau$, and $\Delta \theta$, and can
generate self-consistent and aligned navigation data bits.

Note that {\tt tb2}--{\tt tb6} exhibit a modest amount of quantization and
aliasing noise that makes them easier to detect than would be expected based
on this paper's models, whereas {\tt tb7} is free from such extraneous noise.
{\tt tb7} is also special in that the spoofer exercises control of
$\Delta \theta$, permitting nulling, as described in Section
\ref{sec:nulling-or-blocking}, in the early stages of the attack.  In all
other TEXBAT recordings, the spoofer controlled $\eta$ and $\Delta \tau$ but
left $\Delta \theta$ at an arbitrary constant value ({\tt tb3}, {\tt tb4}, and
{\tt tb6}) or allowed it to ramp consistent with the pull-off rate ({\tt tb2}
and {\tt tb5}).

\subsection{RNL Multipath and Interference Recordings}
This public set of GNSS recordings, from which {\tt wd0}--{\tt wd12} and {\tt
  sm1}--{\tt sm3} are drawn, exhibits mild-to-severe multipath and mild
unintentional jamming ~\cite{wesson2011vsd, pesyna2011tightly}.  Static and
dynamic scenarios are included in both light and dense urban environments
around Austin, TX.  Each 16-bit quantized recording is centered at 1575.42~MHz
with a bandwidth of 10~MHz and at a complex sampling rate of 12.5~Msps.

\subsection{Intentional Jamming Recordings}
Jamming scenarios {\tt jd1}--{\tt jd4} were recorded using a personal privacy
device generating a sawtooth interference waveform with a sweep range of
1550.02--1606.72~MHz and sweep period of 26~$\mu$s (see
~\cite{r_mitch2011cgj}, Table~1, Row~1 and Fig.~8.). This device typifies
low-cost jammers that can be purchased online and easily operated, albeit
illegally. The jamming interference was combined with clean, static receiver
data from a rooftop antenna and re-recorded. Each 16-bit quantized recording
was centered at 1575.42~MHz with a bandwidth of 10~MHz and at a complex
sampling rate of 12.5~Msps.

\subsection{Clean Recordings}
Three recordings ({\tt cd0}, {\tt wd0}, and {\tt wd1}), with negligible
interference beyond thermal noise and multi-access interference, were selected
from the TEXBAT and RNL Multipath and Interference Recordings data sets.
These were all static data sets in quiet RF environments with little
multipath.

\begin{table}[ht]
\centering
\caption{Summary of 27  data recordings used for experimental evaluation.}
\begin{tabular}[c]{clll}
  \toprule
  Type & ID & Description & Duration (s) \\ \midrule

  $H_0$ &{\tt  cd0}& static rooftop & 456 \\
  
  $H_0$ &{\tt wd0}&  static open field & 304 \\
  
  $H_0$ &{\tt wd1}&  static open field  & 298 \\

  $H_1$ &{\tt wd2}&  dynamic deep urban  & 298\\

  $H_1$ &{\tt wd3}&  dynamic deep urban & 456 \\

$H_1$ &{\tt wd4}&  dynamic deep urban  & 292 \\

$H_1$ &{\tt wd5}&  dynamic deep urban  & 292 \\

$H_1$ &{\tt wd6}&  dynamic deep urban  & 460 \\

$H_1$ &{\tt wd7}&  dynamic deep urban  & 246 \\

$H_1$ &{\tt wd8}&  dynamic deep urban  & 278\\

$H_1$ &{\tt wd9}&  dynamic deep urban  & 334 \\

$H_1$ &{\tt wd10}&  dynamic deep urban  & 370 \\

$H_1$ &{\tt wd11}&  dynamic deep urban  & 390 \\

$H_1$ &{\tt wd12}&  dynamic deep urban  & 390 \\

$H_1$ &{\tt sm1}& static urban  & 1700 \\

$H_1$ &{\tt sm2}&  static urban  & 600 \\

$H_1$ &{\tt sm3}&  static urban  & 150 \\

$H_2$ &{\tt tb2} & static time push, $\eta = 10$ dB & 346 \\

$H_2$ &{\tt tb3} & static time push, $\eta = 1.3$ dB & 336 \\

$H_2$ &{\tt tb4} & static pos. push, $\eta = 0.4$ dB & 336 \\

$H_2$ &{\tt tb5} & dynamic pos. push, $\eta = 9.9$ dB   & 304 \\

  $H_2$ &{\tt tb6} & static time push, $\eta = 0.8$ dB  & 308 \\

  $H_2$ &{\tt tb7} & static time push, $\Delta \theta$ control & 468 \\

$H_3$ &{\tt jd1} & PPD, $\eta = 18$ dB & 108 \\

$H_3$ &{\tt jd2} &  PPD, $\eta = 7$ dB  & 58 \\

$H_3$ &{\tt jd3} &  PPD, $\eta = 8$ dB  & 108 \\

$H_3$ &{\tt jd4} &  PPD, $\eta = 2$ dB  & 108 \\
\bottomrule
\end{tabular}
\label{dataDescription}
\end{table}

\subsection{Pre-Processing}
Raw wideband complex samples were first processed by the GRID science-grade
software-defined receiver \cite{lightsey2013demonstration} to produce 100-Hz
complex accumulations free of navigation data modulation at 41
uniformly-spaced taps spanning the range $[-\tau_c,\tau_c]$ around the prompt
tap ($\tau = 0$).  From the resulting data-free 100-Hz accumulations, the
nominal thermal noise deviation $\sigma_{{\rm N}0}$ needed to form $D_k$ in
(\ref{eq:sd}) was estimated by taking a complex-accumulation-wise standard
deviation across multiple channels and multiple correlation tap offsets over a
15 second period in each recording that was substantially free from
interference.  Blocks of 10 100-Hz complex accumulations were then averaged to
create 10-Hz accumulations, at each of the 41 code phase offsets.  These 10-Hz
complex accumulations are those modeled by $\xi_k(\tau)$ in
Fig. \ref{fig:rx-model}.

Received power was calculated directly from the raw wideband (e.g., 25 Msps)
complex samples by filtering to a bandwidth $W_P$ of 2 MHz, as in
Fig. \ref{fig:spectrum}, then averaging the squared modulus of the filtered
samples over 200 ms to produce a 5-Hz time history of received power.  This
was aligned in time with the $\xi_k(\tau)$ and interpolated to produce a
corresponding 10-Hz $P_k$.

\begin{figure}
  \centering
  \includegraphics[width=0.5\textwidth,trim={0.4cm 0.2cm 0.4cm 0.5cm},clip]
  {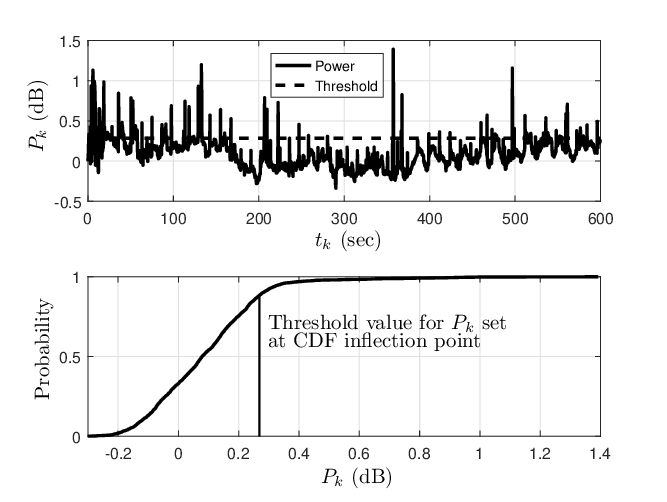}
  \caption{Top: Time history of received power measurements $P_k$ for data set
    {\tt sm2}. Power spikes indicate the presence of low-level unintentional
    jamming.  Bottom: The inflection point of the empirical cumulative
    distribution function (CDF) was taken as the threshold above which
    the empirical data were deemed to be of type $H_3$ (jamming) instead of
    type $H_1$ (multipath), and so excised from the recording.}
  \label{fig:interference}
\end{figure}

Naturally-occurring low-level (less than 2 dB) unintentional jamming was
present in many of the type-$H_1$ (multipath) data recordings taken in urban
settings. Fig.~\ref{fig:interference} shows an example $P_k$ time history
exhibiting such jamming as intermittent spikes.  The short data intervals
containing these spikes were excised from these recordings to ensure they
exhibited only multipath and thermal noise.  Otherwise, the PD detector would
classify the type-$H_1$ recordings as a mixture of multipath and jamming,
which, although true, would complicate an analysis of misclassification using
the empirical data.  As shown in the bottom panel of
Fig.~\ref{fig:interference}, the cumulative distribution of power for each
data set was used to identify the distribution tail, at whose boundary the
excision threshold was set.

\section{Experimental Results}
\label{sec:experiment}
This section documents the experimental performance assessment of the PD
detector, when applying the decision regions shown in Fig. \ref{fig:regions},
against the 27 recordings listed in Table \ref{dataDescription}.  Because {\tt
  tb7} is a special case that violates the PD detector's assumption against
nulling, it will be treated separately in the discussion below.

\begin{figure}
  \centering
  \includegraphics[width=0.5\textwidth,trim={0.3cm 0 0.4cm 0.4cm},clip]
  {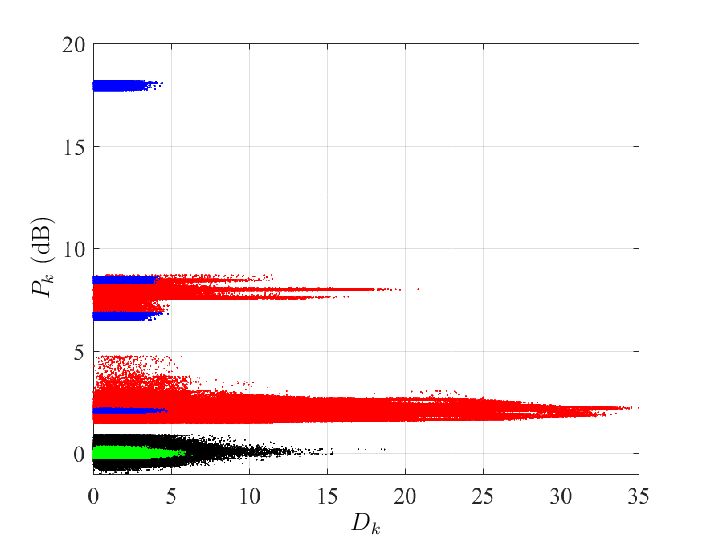}
  \caption{Observations ${z}_k = [D_k, P_k]^T$ for clean (green), multipath
    (black), spoofing (red), and jamming (blue) from the experimental 
    recordings.}
  \label{fig:pincer}
\end{figure}

Observations $z_k$ from the experimental recordings (\textit{sans} {\tt tb7}) are shown
in Fig.~\ref{fig:pincer}.  Clearly, the simulated observations in
Fig.~\ref{fig:sim} agree well with the empirical ones.  The abrupt upper
boundary of the multipath samples is due to the thresholding discussed in
connection with Fig. \ref{fig:interference}.

\begin{figure}[ht]
  \centering
  \includegraphics[width=0.48\textwidth]
  {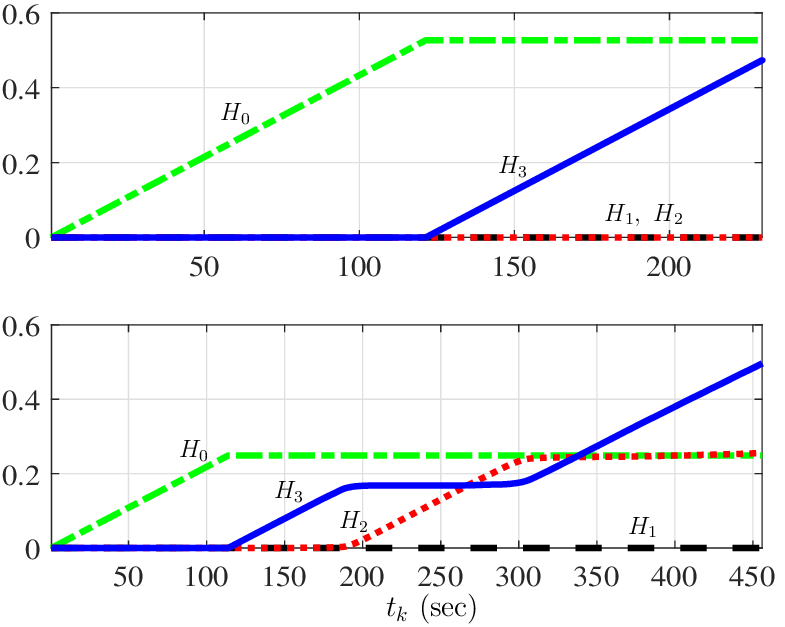}
  \caption{Cumulative time history of decisions $\delta_{\rm B}({z}_k)$ for a
    single receiver tracking channel.  Each trace represents the total number
    of times the corresponding hypothesis was chosen up to time $t_k$,
    normalized so that the final cumulative values sum to one.  Both attacks
    begin at 120 seconds. Top: Jamming scenario {\tt jd3}. Bottom: Spoofing
    scenario {\tt tb4} with the detector applied to a mid-elevation satellite
    signal.}
  \label{fig:csjs}
\end{figure}

Fig. \ref{fig:csjs} shows a single-channel cumulative time history of the PD
detector's decisions for example jamming (top panel) and spoofing (bottom
panel) attack scenarios. In the jamming scenario, the attack is detected
immediately at onset, and continuously declared so thereafter.  In the
spoofing scenario, the attack is detected immediately, but initially
classified as jamming because the spoofer's near-perfect initial
code-phase alignment ($\Delta \tau \approx 0$) causes little distortion (and,
indeed, little harm to the receiver). At about 180 seconds, the spoofer begins
its pull-off, whereupon the increased correlation function distortion reveals
the attack as spoofing.  After $t_k = 300$, the pull-off has proceeded far
enough that correlation function distortion is less pronounced and the attack
is declared to be jamming again.  It is notable that the spoofing attack is
caught despite its low power advantage, which was $\eta = 0.4$ dB as intended,
but $\sim1.5$ dB in effect due to quantization noise in the spoofer.

Table~\ref{tab:misclass} summarizes the PD detector's overall performance
against the experimental data in terms of classification statistics.
Importantly, all instances of spoofing and jamming were flagged as attacks.
However, whereas jamming was always categorized correctly, spoofing was
declared to be jamming $>$4/5 of the time, much higher than the simulated-data
spoofing-to-jamming misclassification rate presented in Table
\ref{tab:misclassSim}.  The difference is that the experimental attacks all
begin initially code-phase aligned ($\Delta \tau \approx 0$) and so do not
initially cause harm or significant distortion.  Moreover, as pull-off
proceeds and $\Delta \theta$ exceeds $\tau_c$, distortion subsides, becoming
negligible beyond $2 \tau_c$, whereupon the spoofing is either classified as
$H_0$ (if low-power) or $H_3$ (if high-power).  Thus, the PD detector is most
effective at recognizing spoofing as such during initial carry-off of the
tracking points.  Conveniently, this is precisely when (1) carry-off-type
spoofing begins to be hazardous, and when (2) civil GNSS spoofing detection
strategies based on cryptographic security codes, as in
\cite{humphreys2011ds}, are least effective.  Thus, the PD detector is nicely
complementary with the methods of \cite{humphreys2011ds}.

A second important result in Table~\ref{tab:misclass} is the low rate of false
spoofing or jamming alarms.  Clean ($H_0$) data produced no false alarms, and
multipath-rich data ($H_1$) produced false alarms only 0.57\% of the time.
Thus, for a multi-channel decision, assuming an urban setting with independent
multipath across tracking channels and $t_{k+1} - t_k = 100$ ms, any subset of
6 from a total of $N \leq 20$ tracking channels would only simultaneously
false alarm on average once every 2.5 years.


\begin{table}[h]
  \centering
  \caption{As Table \ref{tab:misclassSim} but for the PD detector applied to
    the experimental recordings (\textit{sans} {\tt tb7}).}
  \begin{tabular}[c]{ccccc}
    \toprule
     Decision & \multicolumn{4}{c}{True Scenario} \\
    \cmidrule(r){1-1} \cmidrule(r){2-5}
        & $H_0$ & $H_1$ & $H_2$ & $H_3$ \\  \midrule     
     $H_0$ &1 & 0.8717 & 0 & 0 \\
    $H_1$ & 0 & 0.1226 & 0 & 0 \\
    $H_2$ & 0 & 0.0057 & 0.1784 & 0 \\
    $H_3$ & 0 & 0 & 0.8217 & 1 \\
    \bottomrule
  \end{tabular}
\label{tab:misclass}
\end{table}

To test its limits, the PD detector was applied to {\tt tb7}, an especially
subtle attack in which the spoofer carefully controls $\Delta \theta$, effects
authentic signal nulling during the initial stages of the attack, and
maintains an approximately constant measured signal amplitude during pull-off
\cite{texbat_ds7_ds8}.  As explained in Section \ref{sec:nulling-or-blocking},
such an attack would be difficult to mount outside the laboratory.  As might
be expected, the detector's performance was worse for this attack than for the
other spoofing attacks: its decision rates during the attack portion of the
recording were $H_0$: 14\%, $H_1$: 10\%, $H_2$: 70\%, and $H_3$: 6\%.
Nonetheless, the attack was caught on each channel soon after pull-off began.

\section{Conclusions}
\label{sec:conclusion}
We presented the PD detector, a novel low-cost, receiver-autonomous,
readily-implementable GNSS jamming and carry-off spoofing detector.  The
detector traps a would-be attacker between simultaneous monitoring of received
power and complex correlation function distortion.  It amounts to a
multi-hypothesis Bayesian classifier applied to a problem with three unknown
parameters whose prior distributions are informed by the physics of GNSS
signal reception and signal processing, and whose prior probabilities can be
adjusted to reflect the threat environment in which a receiver operates.  In
evaluation against 27 high-quality experimental recordings of attack
and non-attack scenarios, the detector correctly alarmed on all malicious
attacks while maintaining a single-channel false alarm rate below 0.6\%.  For
convenient implementation, the PD detector's decision rule for three different
cost functions, together with all code required to generate
application-tailored decision rules, is available at
https://github.com/navSecurity/P-D-defense.

\section*{Acknowledgements}
Jason Gross's work on this project was supported in part by a West Virginia University Big XII Faculty Fellowship. Todd Humphreys'€™s work on this project has been supported by the National
Science Foundation under Grant No. 1454474 (CAREER) and by the Data-supported
Transportation Operations and Planning Center (DSTOP), a Tier 1 USDOT
University Transportation Center.

\bibliographystyle{IEEEtran} 
\bibliography{pincer}
\end{document}